\pdfoutput=1

\documentclass[12pt]{article}

\usepackage{newtxtext,newtxmath}

\usepackage{graphicx}
\usepackage{float}

\usepackage{xcolor}

\usepackage[letterpaper,margin=1in]{geometry}

\usepackage{amsmath}
\usepackage{bm}

\renewenvironment{abstract}
    {\quotation}
    {\endquotation}

\date{}

\makeatletter
\renewcommand{\fnum@figure}{\textbf{Fig. \thefigure.}}
\renewcommand{\fnum@table}{\textbf{Table \thetable.}}
\long\def\@makecaption#1#2{%
  \vskip\abovecaptionskip
  \sbox\@tempboxa{#1 #2}%
  \ifdim \wd\@tempboxa >\hsize
    #1 #2\par
  \else
    \global\@minipagefalse
    \hb@xt@\hsize{\hfil\box\@tempboxa\hfil}%
  \fi
  \vskip\belowcaptionskip}
\makeatother

\usepackage{scicite}

\usepackage{url}

\newcommand{\TN}{\ensuremath{T_{\mathrm{N}}}}
\newcommand{\TC}{\ensuremath{T_{\mathrm{C}}}}
\newcommand{\thetaK}{\ensuremath{\theta_{\mathrm{K}}}}
\newcommand{\etaK}{\ensuremath{\eta_{\mathrm{K}}}}
\newcommand{\sigmaxy}{\ensuremath{\sigma_{xy}(\omega)}}
\newcommand{\rhoxx}{\ensuremath{\rho_{xx}}}

\newcommand{\rhoAHE}{\ensuremath{\rho^{\mathrm{AHE}}_{yx}}}
\newcommand{\alphaMnTe}{\ensuremath{\alpha}\text{-MnTe}}

\newcommand{\MnthreeSn}{\ensuremath{\mathrm{Mn}_3\mathrm{Sn}}}
\newcommand{\MnthreeNiN}{\ensuremath{\mathrm{Mn}_3\mathrm{NiN}}}
\newcommand{\MnthreeOfour}{\ensuremath{\mathrm{Mn}_3\mathrm{O}_4}}
\newcommand{\CoTaS}{\ensuremath{\mathrm{Co}_{1/3}\mathrm{TaS}_2}}
\newcommand{\microrad}{\ensuremath{\,\mu\mathrm{rad}}}
\newcommand{\nrad}{\ensuremath{\,\mathrm{nrad}}}

\newif\ifembedmainfigs
\embedmainfigstrue

\def\scititle{
    Giant spontaneous Kerr effect and its self-doping dependence in
    altermagnetic MnTe
}
\title{\bfseries \boldmath \scititle}

\author{
    Weitung~Yang$^{1}$,
    Choongjae~Won$^{2}$,
    Cory~D.~Cress$^{3}$,
    Weihang~Lu$^{1}$,\and
    Xiaochen~Fang$^{4}$,
    Shelby~Fields$^{5}$,
    Nicholas~Combs$^{5}$,
    Shaofeng~Han$^{1}$,\and
    Marshall~Zachary~Franklin$^{1}$,
    Steven~P.~Bennett$^{5}$,
    Sang-Wook~Cheong$^{4,6}$,
    Jing~Xia$^{1\ast}$\and
    \small$^{1}$Department of Physics and Astronomy, University of California, Irvine, CA 92697, USA.\and
    \small$^{2}$Laboratory for Pohang Emergent Materials and Max Planck POSTECH Center for Complex Phase Materials,\and\small Pohang University of Science and Technology, Pohang 37673, Korea. \and
    \small$^{3}$Electronics Science and Technology Division, U.S. Naval Research Laboratory,
        Washington, DC 20375, USA.\and
    \small$^{4}$Keck Center for Quantum Magnetism and Department of Physics and
        Astronomy,\and\small Rutgers University, Piscataway, NJ 08854, USA.\and
    \small$^{5}$Materials Science and Technology Division, U.S. Naval Research Laboratory,
        Washington, DC 20375, USA.\and
    \small$^{6}$Center for Quantum Science and Technology and Department of Physics,\and\small
        National TsingHua University, Hsinchu 30013, Taiwan.\and
    \small$^{\ast}$Corresponding author. Email: xia.jing@uci.edu
}

\begin{document}

\maketitle

\begin{center}
\parbox{0.9\textwidth}{\centering
\textbf{Short title:} Giant Kerr Effect in Altermagnetic MnTe}
\end{center}

\vspace{0.3\baselineskip}

\begin{center}
\parbox{0.9\textwidth}{\centering
\textbf{One-sentence summary:}
A giant Kerr effect at a telecommunication wavelength in altermagnetic \alphaMnTe{} tracks
carrier self-doping.}
\end{center}

\vspace{0.5\baselineskip}

\begin{abstract} \bfseries \boldmath
Altermagnetism, a third class of collinear magnetism with spin-split bands and vanishing net
magnetization, has emerged in hexagonal \alphaMnTe{}, a promising platform for ultrafast,
stray-field-free spintronics. Whether MnTe's macroscopic symmetry-breaking signatures reflect
ideal altermagnetic order or are activated by defects remains open.
Here we report giant spontaneous Kerr rotations of up to \mbox{$\pm 1500\microrad$} in
\alphaMnTe{} single crystals at the telecommunication wavelength of \mbox{$1550\,\mathrm{nm}$},
onsetting precisely at the N\'{e}el temperature \mbox{$\TN = 307\,\mathrm{K}$}.
The signal appears in disjoint macroscopic patches whose amplitude tracks the sample
conductivity, and a hole-conducting \alphaMnTe{} film reproduces this response together with an
anomalous Hall effect. Within a single crystal, Kerr-active regions are optically distinct from
their Kerr-silent surroundings in co-registered reflection maps. These observations indicate that
carrier self-doping, rather than ideal altermagnetic order alone, governs the giant
magneto-optical response, and establish telecom-wavelength Kerr imaging as a practical readout
for altermagnetic spintronics.
\end{abstract}

\subsection*{Introduction}

\noindent
Altermagnetism has recently been recognized as a third class of collinear magnetism, distinct from both ferromagnetism and conventional antiferromagnetism~\cite{smejkal2022beyond,smejkal2022emerging,mazin2022editorial}. In altermagnets, two opposite-spin sublattices are related by a rotational crystal symmetry rather than translation or inversion, generating momentum-dependent, nonrelativistic spin-split bands that break time-reversal ($T$) symmetry while preserving a globally compensated, vanishing
net magnetization~\cite{jungwirth2016afm,baltz2018afm}. The combination of ferromagnet-like time-reversal symmetry breaking (TRSB) without stray
fields is widely seen as a promising route to stray-field-free, ultrafast spintronics and optical
electronics beyond what is possible with either ferromagnets or conventional
antiferromagnets~\cite{jungwirth2016afm,baltz2018afm,kriegner2017magnetic,rimmler2025noncollinear}. Among the predicted altermagnets, hexagonal \alphaMnTe{} has become the flagship
compound~\cite{kriegner2017magnetic,krempasky2024altermagnetic,lee2024broken,osumi2024observation}.
Its NiAs-type structure realizes a six-fold spin--crystal symmetry that connects two antiparallel
Mn sublattices via a non-symmorphic rotation combined with a spin flip, producing a $g$-wave
altermagnetic spin-split band structure below the N\'{e}el temperature $\TN \approx 307\,\mathrm{K}$.
Angle-resolved photoemission spectroscopy (ARPES) on both bulk single crystals and epitaxial films has
revealed spin splittings up to ${\sim}0.8\,\mathrm{eV}$ at off-symmetry momenta, in quantitative
agreement with \textit{ab initio} predictions~\cite{krempasky2024altermagnetic,lee2024broken,osumi2024observation}.
A spontaneous anomalous Hall effect (AHE) has been reported in the apparent absence of a
measurable net moment~\cite{gonzalez2023spontaneous}, and altermagnetic domain walls have been
mapped at the $100\,\mathrm{nm}$ scale using nitrogen-vacancy magnetometry and X-ray photoemission
electron microscopy~\cite{amin2024nanoscale}. Current efforts toward MnTe-based altermagnetic
spintronics are focused on practical handles for activating and switching the spin response:
uniaxial strain has been shown to drive a giant spin-splitting effect in this \mbox{$g$}-wave
material~\cite{belashchenko2025giant}, and surface-antisymmetry-group design rules have been identified
for deterministic spin-torque switching of the N\'{e}el vector in altermagnet
films~\cite{belashchenko2026deterministic}.

The magneto-optical Kerr effect
(MOKE)~\cite{faraday1846experimental,kerr1877rotation}, the differential reflection of
oppositely circularly polarized light, has been extensively utilized for over a century for
probing magnetic properties, visualizing magnetic domains, and enabling magneto-optical
technologies~\cite{qiu1999surface,kirilyuk2010ultrafast,mansuripur1995physical}. In
ferromagnets, large MOKE signals arise from the combination of magnetization and spin--orbit
coupling (SOC)~\cite{argyres1955theory}. In coplanar, non-collinear antiferromagnets,
SOC-induced Berry curvature enables MOKE signals as large as $300\microrad$ despite a
vanishingly small net magnetization, as demonstrated in
\MnthreeSn~\cite{higo2018large} and in \MnthreeNiN~\cite{farhang2026temperature}. More recently, large
MOKE signals have been shown to arise in noncoplanar compensated antiferromagnets without SOC,
through real-space scalar spin chirality, as demonstrated in
\CoTaS~\cite{farhang2026topological,feng2020topological}. MOKE also has a long track record of
identifying broken-TRSB phases beyond simple magnets, including the chiral superconducting state
of UPt$_3$~\cite{schemm2014observation} and the orbital Hall effect in the light metal
Ti~\cite{choi2023observation}, motivating its use here as a laboratory-scale, telecom-wavelength
optical probe of altermagnetic order. A natural question is whether, and how, MOKE can be
observed in collinear altermagnets, where spin-split bands and macroscopic TRSB coexist with
vanishing net magnetization~\cite{smejkal2018topological,song2025altermagnets,jungwirth2018multiple}.

A central question remains unresolved: whether the macroscopic TRSB signals in MnTe reflect the
ideal altermagnetic order, or whether they are activated by deviations from stoichiometry and by
defect-induced carriers. Several observations point to the latter. MnTe hosts a tiny, barely
detectable remanent magnetization, recently termed a ``gossamer'' ferromagnetism~\cite{mazin2024origin},
that enables its altermagnetic domains to be trained by external magnetic fields.
Stoichiometry-dependent studies have shown that Mn/Te off-stoichiometry can generate a genuine
ferromagnetic component~\cite{chilcote2024stoichiometry}, and the coexistence of AHE with weak
magnetization has been demonstrated in nominally collinear antiferromagnetic
MnTe~\cite{kluczyk2024coexistence}. Molecular-beam epitaxy (MBE) grown \alphaMnTe{} films have
been reported to exhibit antisite defects, thermal strain from coefficient-of-thermal-expansion
mismatch with the substrate, and a self-limiting surface oxide~\cite{bey2024unexpected,bey2025interface}.
It has also been reported that, across a series of \alphaMnTe{} single crystals with varying
doping levels, \textit{only} those exhibiting a transport activation energy in a narrow
\mbox{$15$}--\mbox{$18\,\mathrm{meV}$} window display the spontaneous AHE: a strong indication that a specific
hole-doped Fermi-level regime is required to activate macroscopic TRSB~\cite{liu2025strain}.

A key microscopic question is the role of self-doping. In practice, as-grown \alphaMnTe{} is
almost always $p$-type: native point defects (primarily Mn vacancies and Mn/Te off-stoichiometry) donate
holes to the valence band and yield low-temperature resistivities in the $10$--$100\,\mathrm{m\Omega{\cdot}m}$
range. This hole self-doping is a generic feature of \alphaMnTe{} samples across essentially
all prior ARPES, transport, and magnetometry
studies~\cite{krempasky2024altermagnetic,lee2024broken,osumi2024observation,gonzalez2023spontaneous,
chilcote2024stoichiometry,kluczyk2024coexistence,liu2025strain}. It has been emphasized
theoretically that real altermagnets always host some degree of SOC, and that the resulting weak
ferromagnetism and TRSB-sensitive responses should be regarded as intrinsic to the material
rather than as artifacts~\cite{cheong2024altermagnetism,cheong2024emergent,cheong2025classification}.
In particular, the ideal altermagnet possesses an effective $T$-symmetry (time reversal combined
with a spin flip) that \textit{forbids} Berry-curvature contributions to the off-diagonal optical
conductivity; these become allowed only when this effective $T$-symmetry is lifted by SOC or by
additional canting~\cite{cheong2024altermagnetism,unconventional2025moke,smejkal2020crystal}.

MOKE, as a direct optical measurement of the off-diagonal conductivity \sigmaxy, is a
particularly sensitive and geometry-complementary probe of this question. The same off-diagonal
tensor also controls the Berry-curvature-induced AHE reported in altermagnetic
MnTe~\cite{gonzalez2023spontaneous}, so transport and optical TRSB responses should respond
together. It was predicted in 2023 that MnTe should exhibit a measurable spontaneous MOKE as a
direct signature of its altermagnetic order~\cite{mazin2023altermagnetism}, but a direct
experimental confirmation has been missing: prior time-resolved MOKE studies on MnTe/InP(111)
films~\cite{gray2024timeresolved} identified magnon and phonon modes but did not resolve a
spontaneous Kerr rotation. MnTe, a collinear centrosymmetric altermagnet, therefore lies in a
fundamentally different regime from the coplanar non-collinear
antiferromagnets~\cite{higo2018large,farhang2026temperature} and the noncoplanar compensated
antiferromagnets~\cite{farhang2026topological,feng2020topological} in which large spontaneous
MOKE has previously been observed, and whether its altermagnetic order, its defect-induced
canting, or both, generate a macroscopic MOKE has remained untested.

In this study, we report the observation of a giant spontaneous MOKE in \alphaMnTe, using a
zero-loop fiber-optic Sagnac interferometer microscope operating at the telecommunication
wavelength of $1550\,\mathrm{nm}$ with a resolution of $10\nrad$. In \alphaMnTe{} single
crystals, we measure Kerr rotations of up to $\pm 1500\microrad$, onsetting precisely at
$\TN = 307\,\mathrm{K}$ and organized into domains of opposite chirality that extend over hundreds
of microns and contain structure resolved down to the micron scale. A modest
$c$-axis field of $0.3\,\mathrm{T}$ reliably trains the domain chirality, and upon warming we
discover a temperature-driven chirality inversion near $150\,\mathrm{K}$. A third, more
resistive bulk crystal shows a reduced spontaneous MOKE of only $\pm 200\microrad$,
suggesting that the Kerr response tracks the sample's hole self-doping. A hole-conducting
\mbox{$58\,\mathrm{nm}$} \alphaMnTe{} film grown by MBE reproduces this behavior in thin-film form,
showing both spontaneous MOKE domains and a hysteretic anomalous Hall effect that vanish
together above \mbox{${\sim}235\,\mathrm{K}$}. Co-registered reflection maps show that the
Kerr-active regions are optically distinct from their Kerr-silent surroundings within a single
crystal. These observations indicate that the spontaneous MOKE in MnTe is not a direct signature
of the ideal altermagnetic order but is activated by carrier self-doping, consistent with the
gossamer-ferromagnetism mechanism~\cite{mazin2024origin} and in line with the
\mbox{$15$}--\mbox{$18\,\mathrm{meV}$} activation-energy window identified for the AHE~\cite{liu2025strain}.

\subsection*{Results}

\subsubsection*{Giant spontaneous Kerr rotations in MnTe bulk crystals}

MOKE detection and imaging were performed using a Sagnac interferometer
microscope~\cite{xia2006modified,xia2006high,xia2008polar} in the polar geometry (Fig.~\ref{fig:1}A),
operating at the telecommunication wavelength of $1550\,\mathrm{nm}$ ($0.80\,\mathrm{eV}$ photon
energy). This technique offers exceptional sensitivity, routinely achieving $10\nrad$ resolution
by using a single-mode optical fiber as both the source and detector for counter-propagating,
time-reversed light beams~\cite{xia2006modified,xia2006high}. The common-path Sagnac geometry
rejects reciprocal polarization effects such as optical birefringence and isolates the
nonreciprocal Kerr response.

High-quality single crystals of hexagonal \alphaMnTe{} (crystals A, B, and C) were grown by the
flux method. A photograph of a representative crystal on
$1\,\mathrm{mm}$ grid paper is shown in the inset of Fig.~\ref{fig:1}A. The NiAs-type structure
features ferromagnetic $ab$-planes stacked antiferromagnetically along the $c$-axis, with
alternating Mn spins that realize the altermagnetic $g$-wave symmetry. The longitudinal
resistivity \rhoxx$(T)$ of crystal A (Fig.~\ref{fig:1}B) exhibits a pronounced peak at
$\TN = 307\,\mathrm{K}$, in agreement with prior
reports~\cite{kriegner2017magnetic,kluczyk2024coexistence}. Crystals A and B show comparable
low-temperature resistivity of ${\sim}10\,\mathrm{m\Omega{\cdot}m}$ at $150\,\mathrm{K}$,
consistent with the generic hole-self-doped regime of as-grown MnTe.

Fig.~\ref{fig:1}C shows the antisymmetrized anomalous Hall resistivity \rhoAHE\mbox{$(B)$} of crystal
A at \mbox{$150\,\mathrm{K}$}. A clear hysteretic loop of $\pm 5\,\mu\Omega{\cdot}\mathrm{m}$ is
observed with the magnetic field $B$ applied along the $c$-axis, consistent with previous
reports on altermagnetic \alphaMnTe{} single crystals and thin
films~\cite{gonzalez2023spontaneous,kluczyk2024coexistence,liu2025strain}. The longitudinal
magnetoresistance and the full Hall signal at \mbox{$150\,\mathrm{K}$} are provided in fig.~\ref{fig:S1}.

Fig.~\ref{fig:2}A--D shows two-dimensional maps of the Kerr rotation \thetaK{} of crystal B
acquired during zero-field warming (ZFW) at $2$, $100$, $200$, and $300\,\mathrm{K}$, following
zero-field cooling (ZFC) from above \TN. We observe giant spontaneous signals of $\pm 1500\microrad$ in
localized domains, five orders of magnitude above our $10\nrad$ noise floor and
comparable in magnitude to the MOKE previously reported in ferromagnetic
metals~\cite{erskine_magneto-optic_1973,kim_surface_1999}. These spontaneous Kerr signals are
five to six times larger than the \mbox{$250$}--\mbox{$300\microrad$} MOKE observed in the non-collinear
antiferromagnet \MnthreeSn~\cite{higo2018large} and noncoplanar antiferromagnet
\CoTaS~\cite{farhang2026topological}, and vanish above \TN, confirming their magnetic origin.

The spontaneous Kerr signals in \mbox{Fig.~\ref{fig:2}A--D}
appear in \textit{disjoint} regions of the sample rather than filling the field of view: large
portions of the crystal surface remain Kerr-silent at the $10\nrad$ level while neighboring
regions exhibit $\pm 1500\microrad$ rotations. This spatial pattern is consistent with the
defect-activation picture developed below, in which the spontaneous MOKE would be switched on only
where the local hole-doping level supports carrier-induced canting.

A first signature of a temperature-driven chirality inversion is visible between Fig.~\ref{fig:2}A
and Fig.~\ref{fig:2}B: several regions that are positive (red) at $2\,\mathrm{K}$ become
negative (blue) by $100\,\mathrm{K}$, at essentially fixed domain boundaries.
\mbox{Fig.~\ref{fig:2}I--L} shows a second ZFW sequence on a different region of the sample,
reproducing the same effect with a distinct nucleation pattern. Fig.~\ref{fig:2}M shows
\thetaK\mbox{$(T)$} traces at three representative locations labeled Points \mbox{1--3.} All three traces
onset at $\TN = 307\,\mathrm{K}$ and track an order-parameter-like temperature dependence.
Point~1 additionally displays a temperature-driven sign inversion of \thetaK{} near
\mbox{$150\,\mathrm{K}$}. The full temperature evolution is provided in figs.~\ref{fig:S2} and \ref{fig:S3}. The onset at
$\TN = 307\,\mathrm{K}$ rules out alternative origins such as antiferromagnetic MnO
($\TN(\mathrm{MnO}) \approx 118\,\mathrm{K}$), metallic Mn clusters ($\TC < 100\,\mathrm{K}$),
\MnthreeOfour{} ($\TC \approx 42\,\mathrm{K}$), or nonmagnetic elemental Te.

Fig.~\ref{fig:3} demonstrates, on crystal A, that a modest $c$-axis field reliably trains the
altermagnetic domain chirality. Fig.~\ref{fig:3}A shows the MOKE map at \mbox{$2\,\mathrm{K}$} and \mbox{$B = -0.3\,\mathrm{T}$}
after field cooling (FC) from \mbox{$350\,\mathrm{K}$}; a uniform negative (blue) Kerr signal develops.
Fig.~\ref{fig:3}B shows that this signal persists as a remanent spontaneous MOKE at $2\,\mathrm{K}$
after the field is removed. Fig.~\ref{fig:3}C further shows that the trained chirality is
preserved upon warming to $100\,\mathrm{K}$. Control measurements across \TN{} (fig.~\ref{fig:S4})
confirm that no MOKE signal develops when the field is applied above \TN, establishing that the
trained state is a genuine below-\TN{} altermagnetic phenomenon.

Upon further warming to $200\,\mathrm{K}$ (Fig.~\ref{fig:3}D), the Kerr sign in several trained
domains \textit{inverts} relative to their $2\,\mathrm{K}$ values, without any change in applied
field. This field-trained inversion confirms the chirality inversion already seen in the ZFC
data and in the Point~1 trace of Fig.~\ref{fig:2}M, now observed across a large ensemble of
co-trained domains. It cannot be explained by conventional domain wall motion alone; it requires
a temperature-dependent sign change of the optical response at essentially fixed N\'{e}el-vector
orientation. A plausible mechanism is a temperature-dependent competition among the
third-order-SOC terms that generate the gossamer canting~\cite{mazin2024origin}. We note that
an AHE sign reversal near $175\,\mathrm{K}$ has also been reported in epitaxial \alphaMnTe{}
films~\cite{bey2024unexpected}; the two sign reversals may share a common microscopic origin.

The cooling field sets both the sign and the extent of the trained state. In a single
\mbox{$200\,\mu\mathrm{m}$} region at \mbox{$150\,\mathrm{K}$}, ZFC leaves both chiralities,
\mbox{$\pm 0.3\,\mathrm{T}$} train most of the region to the corresponding sign, and a smaller
\mbox{$+0.1\,\mathrm{T}$} field trains only part of it (Supplementary Text).

To connect the magneto-optical response directly to the field-driven canting seen in transport
(Fig.~\ref{fig:1}C), we measured \thetaK{} at a fixed position under \mbox{$c$}-axis fields up to
\mbox{$9\,\mathrm{T}$} (Fig.~\ref{fig:3}E). At \mbox{$150\,\mathrm{K}$} the Kerr rotation traces a hysteresis
loop that reverses sign with field, switching abruptly near zero field and approaching
\mbox{$\pm 1400\microrad$} at high field, with a further step near \mbox{$-2\,\mathrm{T}$} on the descending
branch, akin to a metamagnetic transition. The high-field magnitude of this field-driven loop is comparable to the
amplitude of the spontaneous zero-field domains, linking the field-modulated MOKE to the field
sensitivity inferred from the AHE loop of Fig.~\ref{fig:1}C. The right panel of Fig.~\ref{fig:3}E shows a high-resolution
zero-field scan of a \mbox{$200\,\mu\mathrm{m}$} region containing this point, which resolves
\mbox{$\pm 1500\microrad$} domains on a length scale far below that of the full-crystal maps and locates
the measurement point within a single domain.

As the size of the AHE has been reported to vary by orders of magnitude between different
\alphaMnTe{} crystals~\cite{kluczyk2024coexistence,liu2025strain}, we tested whether the
magnitude of the spontaneous MOKE is likewise non-universal. A third crystal, crystal C, grown
under nominally identical conditions to crystals A and B but exhibiting a ten-fold
higher longitudinal resistivity (\mbox{$\rhoxx(150\,\mathrm{K}) \approx 100\,\mathrm{m\Omega{\cdot}m}$}),
shows positive and negative domains with peak Kerr rotations of only \mbox{$\pm 200\microrad$}
(fig.~\ref{fig:S5}), approximately an order of magnitude smaller than in crystals A and B, while still
four orders of magnitude larger than our \mbox{$50\nrad$} spatially averaged noise floor.

\subsubsection*{Spontaneous MOKE and anomalous Hall effect in a conducting MnTe film}

Fig.~\ref{fig:4} shows that the behavior of the bulk crystals is reproduced in thin-film
form. A \mbox{$58\,\mathrm{nm}$} \alphaMnTe{} film was grown by MBE on a lattice-matched
InP(111)A substrate (Materials and Methods). The film is hole-conducting, with a two-terminal
resistance in the \mbox{$100\,\mathrm{k\Omega}$} range.

Fig.~\ref{fig:4}A shows \thetaK\mbox{$(T)$} measured at the fixed position marked by a black circle in
\mbox{Figs.~\ref{fig:4}D--F}, together with the anomalous Hall resistivity \rhoAHE{} extracted
from the field sweeps of \mbox{Figs.~\ref{fig:4}B and C}. Both quantities fall to zero above
\mbox{${\sim}235\,\mathrm{K}$}. At \mbox{$200\,\mathrm{K}$} the Hall loop is clearly hysteretic
with a saturated amplitude of \mbox{${\sim}4.4\,\mu\Omega{\cdot}\mathrm{m}$}, the thin-film
counterpart of the bulk loop of Fig.~\ref{fig:1}C; by \mbox{$250\,\mathrm{K}$} it has closed.
\mbox{Figs.~\ref{fig:4}D--F} show zero-field MOKE maps at \mbox{$50$}, \mbox{$150$}, and
\mbox{$250\,\mathrm{K}$}: at \mbox{$50\,\mathrm{K}$} the film carries well-resolved positive and
negative Kerr domains reaching \mbox{$\pm 200\microrad$}, which weaken by
\mbox{$150\,\mathrm{K}$} and are absent at \mbox{$250\,\mathrm{K}$}.

The principal connection between this film and the bulk crystals is that MOKE and AHE appear
together and vanish together at the same temperature, as expected for two limits of the same
off-diagonal conductivity \sigmaxy. The film thus reproduces in thin-film form the joint optical
and transport signature of the self-doped bulk crystals. We do not place the film on the same
resistivity axis as the bulk crystals: its two-terminal geometry and its epitaxial strain and
domain structure differ from those of a cleaved bulk face, so its Kerr amplitude is not directly
comparable to theirs.

The expected response of the ideal material follows first from symmetry. In the strict
nonrelativistic limit, the collinear altermagnetic state of \alphaMnTe{} possesses an effective
\mbox{$T$}-symmetry, time reversal combined with a spin flip, that forbids any Berry-curvature
contribution to \sigmaxy{} at every photon
energy~\cite{unconventional2025moke,smejkal2020crystal}; the surviving quantum-metric contribution
is expected to lie below our sensitivity (Supplementary Text). Relativistic SOC lifts this
symmetry, but in \alphaMnTe{} the leading SOC-induced canting enters only at third
order~\cite{mazin2024origin}, so the magneto-optical response of the pristine, stoichiometric
material is suppressed on symmetry grounds alone, independently of any numerical estimate.

The first-principles MOKE spectrum of ideal, undoped altermagnetic \alphaMnTe{}
(fig.~\ref{fig:S6}), computed following the previously established
methodology~\cite{mazin2023altermagnetism}, corroborates this at our probe energy. It yields Kerr
rotations and ellipticities of up to \mbox{$\pm 1500\microrad$}, but that magnitude is reached
near \mbox{$2.3\,\mathrm{eV}$}; at the \mbox{$0.80\,\mathrm{eV}$} energy probed here the
predicted rotation of the ideal state falls to near zero. Because the optical calculation is not
fully \mbox{$k$}-mesh converged below approximately \mbox{$1\,\mathrm{eV}$}, we treat this value
as corroborative rather than load-bearing; the symmetry argument above does not depend on it.
Two conclusions follow. The altermagnetic band structure is intrinsically capable of supporting
Kerr rotations of order \mbox{$1000\microrad$} once its effective \mbox{$T$}-symmetry is broken,
so the magnitude we measure is not anomalous for this material. At the same time, neither the
symmetry argument nor the undoped calculation accounts for the \mbox{$\pm 1500\microrad$} signal
we observe at \mbox{$0.80\,\mathrm{eV}$} in the hole-conducting bulk crystals, which is consistent
with activation by carrier self-doping.

Among the bulk crystals, which were grown under nominally identical conditions and measured
on cleaved \mbox{$c$}-axis faces, the spontaneous MOKE amplitude tracks the sample conductivity:
crystals A and B (\mbox{${\sim}10\,\mathrm{m\Omega{\cdot}m}$}, \mbox{$\pm 1500\microrad$})
against the more resistive crystal C (\mbox{${\sim}100\,\mathrm{m\Omega{\cdot}m}$},
\mbox{$\pm 200\microrad$}). This is the ordering expected if the macroscopic TRSB response is
switched on by hole self-doping. The conducting film enters the argument separately, through the
joint appearance of MOKE and AHE with a common onset temperature rather than through its
amplitude.

\subsection*{Discussion}

Taken together, the bulk-crystal conductivity contrast and the thin-film corroboration
indicate that the spontaneous MOKE in \alphaMnTe{} is activated by carrier self-doping rather than
arising directly from the ideal altermagnetic order. This optical-conductivity ordering is
consistent with a recent transport observation: a survey~\cite{liu2025strain} of \alphaMnTe{} single crystals with varying
doping and strain conditions has shown that \textit{all} samples exhibiting the spontaneous AHE
cluster in a narrow transport activation-energy window of $15$--$18\,\mathrm{meV}$, while
samples outside this window show no AHE. The AHE and MOKE are two measurements of the same
off-diagonal conductivity tensor \sigmaxy{} (the former at $\omega = 0$ and the latter at
$0.80\,\mathrm{eV}$), and both our optical data and the transport survey~\cite{liu2025strain}
point to the same conclusion: macroscopic TRSB in MnTe is switched on only when the Fermi
level sits in a specific hole-doped window.

The stronger constraint comes from within a single crystal. Beyond the global,
sample-averaged correlation above, the co-registered reflection channel
provides spatially resolved evidence that Kerr-active regions are optically distinct, consistent
with locally defect-rich, self-doped material. The instrument records the
reflected optical power \mbox{$P_0$} at every pixel simultaneously with \thetaK, so the magneto-optical
and optical channels are intrinsically co-registered and require no separate measurement or image
registration. \mbox{Fig.~\ref{fig:2}E--H} presents the \mbox{$P_0$} maps acquired together with the
Kerr images of \mbox{Fig.~\ref{fig:2}A--D}, and \mbox{figs.~\ref{fig:S7}--\ref{fig:S9}} present these
and the corresponding maps for Fig.~\ref{fig:3} and fig.~\ref{fig:S5} at larger scale. In the two conducting crystals,
the regions carrying \mbox{$\pm 1500\microrad$} Kerr rotation coincide with localized maxima of the
reflected power, reaching \mbox{${\sim}20\,\mu\mathrm{W}$} against a \mbox{$5$--$10\,\mu\mathrm{W}$} background
elsewhere on the same crystal face, an enhancement of roughly a factor of two. The correlation is
one-directional: every Kerr-active region is optically bright, while not every bright region
carries a Kerr signal, as expected if elevated local doping is necessary but not sufficient for
the carrier-induced canting to develop. This also resolves the apparent tension between a global
transport measurement and a locally varying optical signal: the bulk resistivity is dominated by
the most conductive percolating path through the crystal, whereas \thetaK{} reports the local
carrier and canting state pixel by pixel, so a globally conducting crystal can still contain
extended Kerr-silent regions of lower local doping.

The \mbox{$P_0$} contrast is temperature-independent and persists unchanged up to \mbox{$300\,\mathrm{K}$},
where the Kerr signal has vanished entirely (\mbox{Fig.~\ref{fig:2}D,H} and \mbox{fig.~\ref{fig:S8}E,J}). The
optical feature is therefore a static property of the material rather than a consequence of the
magnetic state, and it cannot be an artifact of the Kerr signal itself. The hierarchy of the
optical contrast tracks the ordering of the MOKE amplitude across the three bulk crystals: in the more
resistive crystal C, whose peak Kerr rotation is an order of magnitude smaller, the
regions of elevated reflected power are correspondingly less pronounced (fig.~\ref{fig:S9}). We note that the reflected power in
the confocal geometry of the Sagnac microscope is not an absolute reflectivity, and that a single intensity channel cannot
be inverted to obtain the optical constants; we therefore do not assign a specific microscopic
optical model.

The probe energy is well matched to this comparison. At \mbox{$0.80\,\mathrm{eV}$} the photon
energy lies below the semiconducting gap of \alphaMnTe{}, so intrinsic interband absorption is
inactive and the ideal, stoichiometric material is transparent at the probe wavelength; the dark,
reflective appearance of the crystal in the photograph of Fig.~\ref{fig:1}A is an above-gap
property at visible wavelengths. Because \mbox{$0.80\,\mathrm{eV}$} lies below the interband gap,
defect- or carrier-related subgap absorption is one plausible contributor to the local optical
contrast, for instance through Mn-vacancy in-gap states, which would raise the local extinction
coefficient at the probe energy. Two limits on this reading should be stated. Transparent
does not mean non-reflecting: the calculated below-gap dielectric function of \alphaMnTe{}
(fig.~\ref{fig:S6}C) implies a normal-incidence Fresnel reflection of roughly one quarter of the
incident power at the front surface of the ideal material, so the observed factor-of-two enhancement sits on a large baseline and cannot be assigned
to absorption alone from a single intensity channel. And because the collected power in a confocal
geometry also depends on local surface flatness and focus, we rely on the correspondence between
the reflection ordering and the MOKE ordering across the bulk crystals, rather than on the contrast
within one image alone, in associating the optically bright regions with elevated local doping.

A free-carrier Drude mechanism is unlikely to dominate, since at the relevant carrier
densities the probe energy lies well above the screened plasma edge, where additional carriers
reduce rather than enhance the reflectance. We further note that energy-dispersive x-ray
spectroscopy (EDS) cannot resolve the relevant compositional variation: the hole densities of order
\mbox{$10^{19}\,\mathrm{cm^{-3}}$} reported for as-grown
\alphaMnTe~\cite{gonzalez2023spontaneous,chilcote2024stoichiometry,kluczyk2024coexistence}
correspond to Mn-vacancy fractions of order \mbox{$10^{-4}$} in the Mn:Te
ratio, one to two orders of magnitude below the \mbox{${\sim}1\%$} relative precision of EDS
quantification. Taken together, these observations provide spatially resolved evidence that the
Kerr-active regions are optically distinct from their surroundings, consistent with locally
defect-rich, self-doped material, and they extend the global \mbox{resistivity--MOKE} correlation
between the bulk crystals down to the micron scale within a single crystal.

The Kerr response of the conducting film is likewise non-uniform and co-registered with a
static optical contrast (fig.~\ref{fig:S10}). The \mbox{$P_0$} pattern of the film is unchanged
between \mbox{$50$} and \mbox{$250\,\mathrm{K}$}, over which range the Kerr signal decays to zero,
so it is not a consequence of the magnetic state. We do not, however, read the film's spatial
pattern as evidence of the same origin as in the bulk, because the thin-film geometry admits
additional contributions to the reflected power that the cleaved bulk faces do not (Supplementary
Text). The spatial argument therefore rests on the bulk crystals.

A competing interpretation has recently been proposed~\cite{zhao2026emergent}, based on
first-principles calculations, that surface states within the bulk gap acquire a
ferromagnet-like spin polarization and dominate the AHE in \alphaMnTe{} thin films at
experimentally relevant Fermi energies. Our observations bear on this scenario. The
spontaneous Kerr response of the bulk crystals appears in disjoint patches that correlate with
local optical contrast, rather than uniformly across the sample surface. The optical contrast with which these patches coincide is itself
temperature-independent and persists unchanged to \mbox{$300\,\mathrm{K}$}, where the Kerr signal
has vanished (\mbox{Fig.~\ref{fig:2}D,H}), so the Kerr-active regions are set by a static, spatially
inhomogeneous property of the material rather than by an electronic feature common to the entire
surface. The same locations remain Kerr-active after zero-field cooling and after cooling in
\mbox{$+0.1$}, \mbox{$-0.3$}, and \mbox{$+0.3\,\mathrm{T}$}, with the cooling field setting only
the chirality (\mbox{figs.~\ref{fig:S11} and \ref{fig:S12}}). We therefore regard bulk carrier self-doping and the
attendant gossamer-canting channel~\cite{mazin2024origin} as the more likely driver of the giant
MOKE in our samples, while noting that MOKE and AHE probe different-frequency limits of the
off-diagonal conductivity.

A natural microscopic interpretation is provided by the gossamer-ferromagnetism
scenario~\cite{mazin2024origin}. In the purely collinear, defect-free altermagnetic state,
\alphaMnTe{} possesses an effective \mbox{$T$}-symmetry that enforces zero Berry-curvature contribution
to
\sigmaxy~\cite{cheong2024altermagnetism,unconventional2025moke,smejkal2020crystal}; any MOKE
must then arise from the much weaker quantum-metric contribution, expected to be below our
current sensitivity. Self-doping introduces itinerant holes into the narrow valence band at the
top of the Mn $d$-manifold, which lower their kinetic energy by driving a small
third-order-SOC canting of the Mn moments out of perfect collinearity; the canted moment grows
approximately linearly with carrier density~\cite{mazin2024origin}. This canting breaks the
effective $T$-symmetry, switches on the full Berry-curvature contribution of the altermagnetic
bands to \sigmaxy, and produces a large MOKE whose magnitude is set not by the tiny canting
angle itself but by the altermagnetic band structure it unblocks. This framework accounts for
the ordering of MOKE amplitudes with carrier density; the onset of the signal precisely
at \TN; and the trainability of the domain chirality by modest $c$-axis fields. Because the
canted moment grows monotonically with carrier density, the framework also implies a doping
threshold below which no macroscopic response develops. It does not, however, account for the
upper edge of the \mbox{$15$}--\mbox{$18\,\mathrm{meV}$} activation-energy window reported for the
AHE~\cite{liu2025strain}, above which the response is again absent; a canting linear in carrier
density has no such upper bound, and an additional ingredient is required.

The defect-activated picture indicated here is entirely consistent with the broader
view~\cite{cheong2024altermagnetism,cheong2024emergent,cheong2025classification} that real
altermagnets always host some degree of SOC, and that the resulting weak ferromagnetism, AHE,
and Kerr effects should be regarded as intrinsic material responses. \alphaMnTe{} belongs to the
``strong'' altermagnet category at the band-structure level: its $g$-wave spin splitting is
nonrelativistic and observable directly by ARPES even in pristine
samples~\cite{krempasky2024altermagnetic,lee2024broken,osumi2024observation}. The MOKE response
of MnTe, however, sits one level deeper in the symmetry hierarchy, at the level of the
off-diagonal optical conductivity. Our data show that at this level SOC is essential: the ideal
nonrelativistic altermagnet is magneto-optically silent at the Berry-curvature level, and a
finite MOKE requires the simultaneous action of SOC and carrier-induced canting of the Mn
moments.

It is instructive to place our MnTe result in the broader landscape of spontaneous MOKE in
compensated magnets. In the coplanar non-collinear antiferromagnets \MnthreeSn~\cite{higo2018large}
and \MnthreeNiN~\cite{farhang2026temperature}, large spontaneous Kerr rotations of ${\sim}300\microrad$
arise from SOC-induced Berry curvature in a triangular spin structure; in the noncoplanar
antiferromagnet \CoTaS~\cite{farhang2026topological}, a Kerr rotation of $250\microrad$ arises from real-space scalar spin chirality without invoking SOC. The MOKE in \alphaMnTe{} belongs to
a distinct third category: a collinear altermagnet, in which the ideal undoped calculation yields near-zero Kerr rotation at the \mbox{$0.80\,\mathrm{eV}$} probe energy, and in which a giant signal emerges only when defect-induced carriers activate a small SOC-driven canting.

Our results carry several implications. First, they reconcile the seemingly contradictory
observations of large spontaneous AHE~\cite{gonzalez2023spontaneous}, coexistence of AHE with
weak magnetization~\cite{kluczyk2024coexistence}, stoichiometry-induced
ferromagnetism~\cite{chilcote2024stoichiometry}, and strain-tuned AHE with
activation-energy-dependent hysteresis~\cite{liu2025strain}: all are different manifestations of
the same underlying carrier-activated gossamer-canting channel. Second, they caution against
using the magnitude of macroscopic TRSB signals as a direct measure of altermagnetic order,
since in pristine samples the ideal altermagnetic order may be fully present and visible to
ARPES~\cite{krempasky2024altermagnetic,lee2024broken,osumi2024observation} or
x-ray magnetic circular dichroism (XMCD)~\cite{hariki2024xray}, while producing a vanishingly small macroscopic response.

Testing this picture directly will require material that is not presently available. To our
knowledge no near-stoichiometric \alphaMnTe{} film or single crystal has been reported: the ARPES,
transport, and magnetometry studies of this compound, including the present
one~\cite{krempasky2024altermagnetic,lee2024broken,osumi2024observation,gonzalez2023spontaneous,
chilcote2024stoichiometry,kluczyk2024coexistence,liu2025strain}, have been performed on
hole-self-doped material, and identifying growth conditions that yield a near-stoichiometric
sample would constitute a significant advance in its own right. The carrier-activation picture
makes a sharp prediction for such a sample: a strongly suppressed spontaneous MOKE and AHE in a
material whose \mbox{$g$}-wave altermagnetic band structure remains fully developed and directly
observable by ARPES. That measurement would be the decisive test of the mechanism indicated here.
The conducting film corroborates the joint magneto-optical and transport response, but a
near-stoichiometric sample suitable for the decisive comparison is not presently available. We
therefore present carrier self-doping as the interpretation indicated by the combined evidence
rather than as conclusively demonstrated.

In summary, we have observed a giant spontaneous MOKE of \mbox{$\pm 1500\microrad$} in altermagnetic
\alphaMnTe{} single crystals using a Sagnac interferometer at the telecommunication wavelength
of \mbox{$1550\,\mathrm{nm}$} with a resolution of \mbox{$10\nrad$}. The Kerr rotation onsets precisely at
$\TN = 307\,\mathrm{K}$, organizes into altermagnetic domains that extend over hundreds of microns
and contain structure resolved down to the micron scale, whose chirality is
trainable by a $0.3\,\mathrm{T}$ field, and exhibits a temperature-driven chirality inversion
near $150\,\mathrm{K}$. A hole-conducting \mbox{$58\,\mathrm{nm}$} \alphaMnTe/InP(111) film grown by MBE reproduces
the same response, with \mbox{$\pm 200\microrad$} domains and a hysteretic AHE that vanish
together above \mbox{$\sim 235\,\mathrm{K}$}. A third bulk crystal, ten times more resistive than
crystals A and B, shows a spontaneous MOKE an order of magnitude smaller, \mbox{$\pm 200\microrad$}.
This conductivity ordering among the bulk crystals is consistent with the
\mbox{$15$}--\mbox{$18\,\mathrm{meV}$} activation-energy window for the AHE in
MnTe~\cite{liu2025strain} and supports a carrier-self-doping origin for macroscopic TRSB, rather than
the ideal altermagnetic order itself. For the
integrated-photonics community, the entire demonstration is carried out at \mbox{$1550\,\mathrm{nm}$},
the workhorse wavelength of the C-band fiber-optic telecommunications infrastructure.
\alphaMnTe, with \TN{} above room temperature, a semiconducting bandgap compatible with
integrated-photonic platforms, a lattice match to the InP(111) substrate, and a giant Kerr
response directly in the C-band, is effectively a ready-made material for the existing
telecommunications fiber infrastructure and for monolithic integration with established silicon-
and InP-photonics platforms.

\vspace{0.5\baselineskip}
\noindent\textbf{Note added.}
During the preparation and review of this manuscript we became aware of two related studies. An
independent, concurrent study reported scanning Kerr imaging of macroscopic altermagnetic domains
in bulk \alphaMnTe{} at telecom infrared wavelength, with large Kerr rotations that do not scale
with the tiny bulk magnetization~\cite{watanabe2026magnetooptical}. That decoupling is consistent
with the mechanism indicated here, in which the Kerr amplitude is set by the altermagnetic band
structure unblocked by carrier-induced canting rather than by the size of the canted moment. A
subsequent polarized-neutron study showed that the altermagnetic order and its coupled,
spin-orbit-induced weak ferromagnetic moment can both be switched by millitesla-scale field
cooling~\cite{liu2026observation}, thereby corroborating, with a bulk magnetic probe, the coupling between
the N\'{e}el vector and a small canted moment that underlies the domain training of
Fig.~\ref{fig:3}.

\subsection*{Materials and Methods}

\subsubsection*{Crystal growth}

High-quality bulk single crystals of hexagonal \alphaMnTe{} (crystals A, B, and C) were grown
by the flux method from stoichiometric mixtures of high-purity Mn (99.99\%) and Te
(99.9999\%) sealed in evacuated quartz ampoules. The resulting crystals were oriented by Laue
X-ray diffraction and cleaved along $c$-axis faces to expose atomically flat surfaces for
optical measurements. Crystals A and B exhibit low-temperature longitudinal resistivity of
$\rhoxx(150\,\mathrm{K}) \approx 10\,\mathrm{m\Omega{\cdot}m}$, in the typical hole-self-doped
regime of as-grown \alphaMnTe. Crystal C, grown under nominally identical conditions,
is more resistive, with $\rhoxx(150\,\mathrm{K}) \approx 100\,\mathrm{m\Omega{\cdot}m}$.

\subsubsection*{Thin-film deposition}

The \mbox{$58\,\mathrm{nm}$} \alphaMnTe{} thin film was grown by molecular-beam epitaxy on a
lattice-matched semi-insulating InP(111)A substrate. \alphaMnTe{} was synthesized by
co-evaporating high-purity elemental Mn and Te from Knudsen cells at a Mn:Te
beam-equivalent-pressure ratio of approximately 1:5, with the substrate temperature maintained
at $300$--$350\,{}^{\circ}\mathrm{C}$. The film was measured as grown, without lithographic patterning; electrical contacts were
made by indium soldering. It is hole-conducting, with a two-terminal resistance in the
\mbox{$100\,\mathrm{k\Omega}$} range.

\subsubsection*{Transport measurements}

DC transport measurements were performed in a commercial Physical Property Measurement System (PPMS) with a resistance bridge. An
electric current was applied along the $ab$-plane of the crystal, and the magnetic field was
applied along the $c$-axis. The anomalous Hall component was extracted by antisymmetrizing
$\rho_{yx}(B)$ and subtracting the linear ordinary Hall background.

\subsubsection*{Sagnac MOKE measurements}

The MOKE measurements were performed using a zero-loop fiber-optic Sagnac
interferometer~\cite{xia2006modified,xia2006high} operating at the telecommunication wavelength
of $1550\,\mathrm{nm}$ ($0.80\,\mathrm{eV}$ photon energy). For MOKE imaging, we used a
scanning Sagnac microscope with $2\,\mu\mathrm{m}$ lateral spatial resolution installed inside a
cryostat with $1.8\,\mathrm{K}$ base temperature. The shot-noise-limited instrument sensitivity
is \mbox{${\sim}10\nrad$}. Reflected-power maps were acquired simultaneously with Kerr maps and used
to verify correct alignment on the sample surface.
The common-path geometry of the interferometer rejects reciprocal polarization effects,
including reciprocal birefringence and optical rotation, preventing them from contributing to the
measured nonreciprocal signal; a rejection of 55 dB for optical anisotropy has been demonstrated
for this instrument~\cite{farhang2026topological}.

\subsubsection*{First-principles MOKE calculations}

We used the results of the density functional theory (DFT) calculations published in
Ref.~\cite{liebman-pelaez_strain_2026}. It should be noted that the \mbox{$k$}-mesh convergence of
optical calculations rapidly deteriorates at small photon energies and the calculations performed
there were not fully converged below \mbox{${\sim}1\,\mathrm{eV}$}; our $0.80\,\mathrm{eV}$ measurement
lies near this boundary, and the DFT--experiment comparison should therefore be interpreted
semi-quantitatively at the measurement wavelength.

\clearpage

\bibliography{references}
\bibliographystyle{sciencemag-allauthors}

\section*{Acknowledgments}
We thank Igor Mazin for providing the full calculational data used in
Ref.~\cite{liebman-pelaez_strain_2026}.

\paragraph*{Funding:}
This project was supported by NSF award DMR-2419425 and the Gordon and Betty Moore Foundation
EPiQS Initiative, Grant \# GBMF10276 (J.X.). The work at Rutgers University was supported by
the DOE under Grant No.\ DE-FG02-07ER46382 (S.W.C.). S.W.C. acknowledges the support from the Yushan Scholar Program under the Ministry of
Education of Taiwan. The work at U.S. Naval Research Laboratory was supported by the Office of
Naval Research 6.1 Base Funding at the U.S. Naval Research Laboratory (S.P.B.). The work at Pohang
University of Science and Technology was supported by the National Research Foundation of Korea
(NRF), Ministry of Science and ICT (RS-2022-NR068223) (C.W.).

\paragraph*{Author contributions:}\mbox{}\\
Conceptualization: JX, SWC, SPB\\
Methodology: JX\\
Investigation: WY, WL, SH, MZF, JX\\
Resources: CW, CDC, XF, SF, NC, SPB, SWC\\
Supervision: JX, SWC, SPB\\
Writing \mbox{--} original draft: JX\\
Writing \mbox{--} review and editing: WY, CW, CDC, WL, XF, SF, NC, SH, MZF, SPB, SWC, JX

\paragraph*{Competing interests:}
The authors declare that they have no competing interests.

\paragraph*{Data, Code, and Materials Availability:}
All data and code needed to evaluate and reproduce the results in the paper are present in
the paper and/or the Supplementary Materials. Source data for all main-text and supplementary
figures are additionally deposited in a figshare repository under DOI
\url{https://doi.org/10.6084/m9.figshare.32086938}, which is publicly accessible and requires
no access code. No new code was generated in this study; the first-principles results reproduced
in \mbox{fig.~\ref{fig:S6}} were computed in Ref.~\cite{liebman-pelaez_strain_2026}.
This study generated new materials. The bulk \alphaMnTe{} single crystals A, B, and C were grown
by the flux method, and the 58 nm \alphaMnTe{}/InP(111)A film was grown by molecular-beam epitaxy,
under the growth conditions given in Materials and Methods. The \alphaMnTe{} single crystals A, B,
and C and the 58 nm \alphaMnTe{}/InP(111)A film used in this study can be provided by Jing Xia,
subject to availability. Requests should be submitted to Jing Xia (xia.jing@uci.edu).

\subsection*{Supplementary materials}
Supplementary Text\\
Figs. S1 to S12

\begingroup
\ifembedmainfigs\else\renewcommand{\includegraphics}[2][]{}\fi

\begin{figure}
\centering
\includegraphics[width=0.95\textwidth]{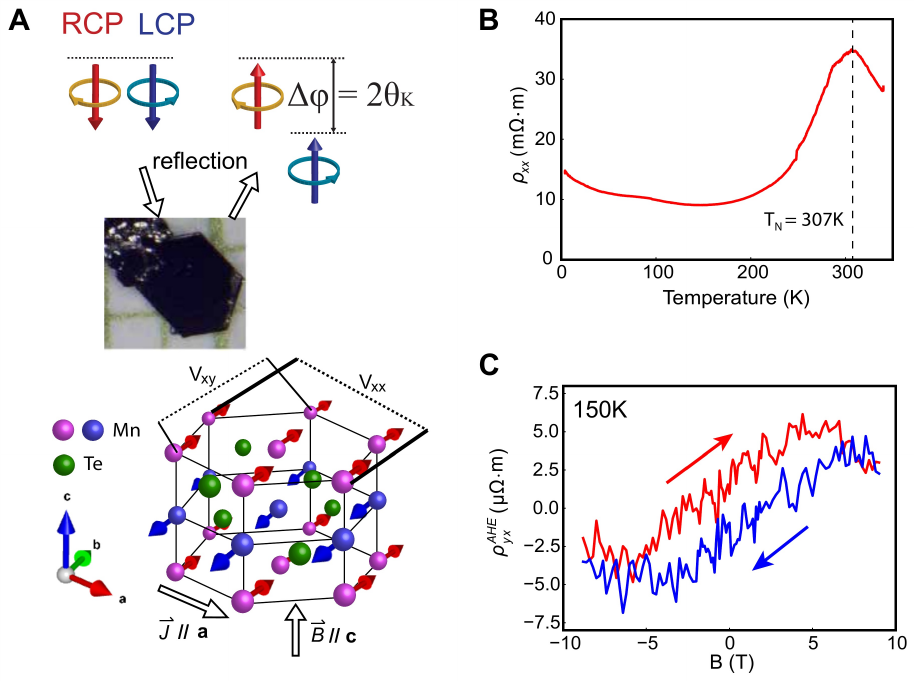}
\caption{\textbf{Anomalous Hall effect (AHE) in MnTe.}
    (\textbf{A}) Experimental setup for optical and transport measurements. RCP and LCP denote
    right- and left-circularly polarized light, which acquire a relative phase shift
    \mbox{$\Delta\varphi = 2\thetaK$} upon reflection. Inset: photograph of
    the \alphaMnTe{} single crystal on \mbox{$1\,\mathrm{mm}$} grid paper; arrows in the hexagonal
    lattice indicate the alternating spin arrangement of the altermagnetic \mbox{$g$}-wave order.
    (\textbf{B}) Longitudinal resistivity \rhoxx{} versus temperature, showing a pronounced peak
    at the N\'{e}el temperature \mbox{$\TN = 307\,\mathrm{K}$}.
    (\textbf{C}) Anomalous Hall resistivity \rhoAHE{} measured at \mbox{$150\,\mathrm{K}$} (where the
    AHE signal is maximized), showing a hysteretic \mbox{$\pm 5\,\mu\Omega{\cdot}\mathrm{m}$} loop with
    \mbox{$B$} along the \mbox{$c$}-axis.}
\label{fig:1}
\end{figure}

\begin{figure}
\centering
\includegraphics[width=0.88\textwidth]{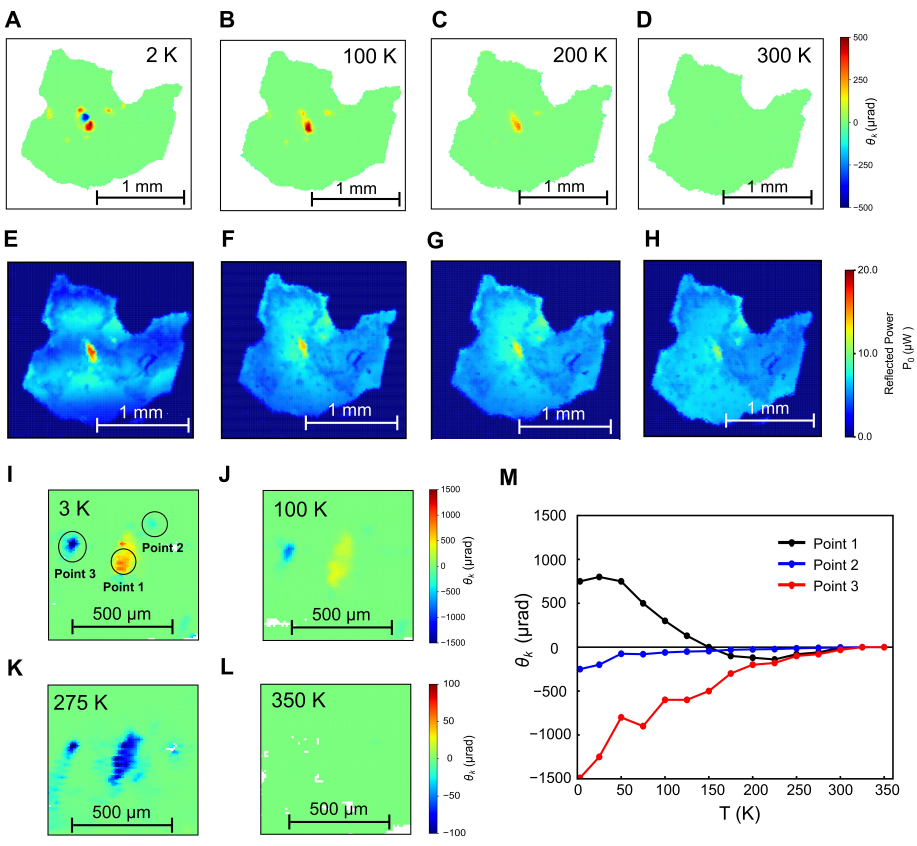}
\caption{\textbf{Spontaneous MOKE domains following zero-field cooling (ZFC) in MnTe bulk
    crystal B.}
    \mbox{(\textbf{A--D})} MOKE images acquired during zero-field warming (ZFW) at \mbox{$2$},
    \mbox{$100$}, \mbox{$200$}, and \mbox{$300\,\mathrm{K}$} on a \mbox{$\pm 500\microrad$} color
    scale, which the domains saturate; their peak amplitude of \mbox{$\pm 1500\microrad$} is given by
    the \thetaK\mbox{$(T)$} traces in panel (M).
    \mbox{(\textbf{E--H})} Reflected optical power \mbox{$P_0$} recorded in the same scan as
    \mbox{(A--D)}, and therefore co-registered with it, on a \mbox{$0$--$20\,\mu\mathrm{W}$} scale.
    \mbox{(\textbf{I--L})} A second ZFW sequence (without training) at \mbox{$3$}, \mbox{$100$},
    \mbox{$275$}, and \mbox{$350\,\mathrm{K}$}; (K) and (L) use a finer
    \mbox{$\pm 100\microrad$} scale. The full temperature evolution is provided in
    figs.~\ref{fig:S2} and \ref{fig:S3}.
    White pixels mark insufficient reflected power; images were registered to fixed surface
    features to correct for thermal drift.
    (\textbf{M}) Kerr rotation \thetaK{} at Points 1, 2, and 3 marked in panel (I), showing onset at
    \mbox{$\TN = 307\,\mathrm{K}$}. Point 1 exhibits a temperature-driven sign reversal near
    \mbox{$150\,\mathrm{K}$}, the chirality inversion also seen on trained domains in
    Fig.~\ref{fig:3}D.}
\label{fig:2}
\end{figure}

\begin{figure}
\centering
\includegraphics[width=0.80\textwidth]{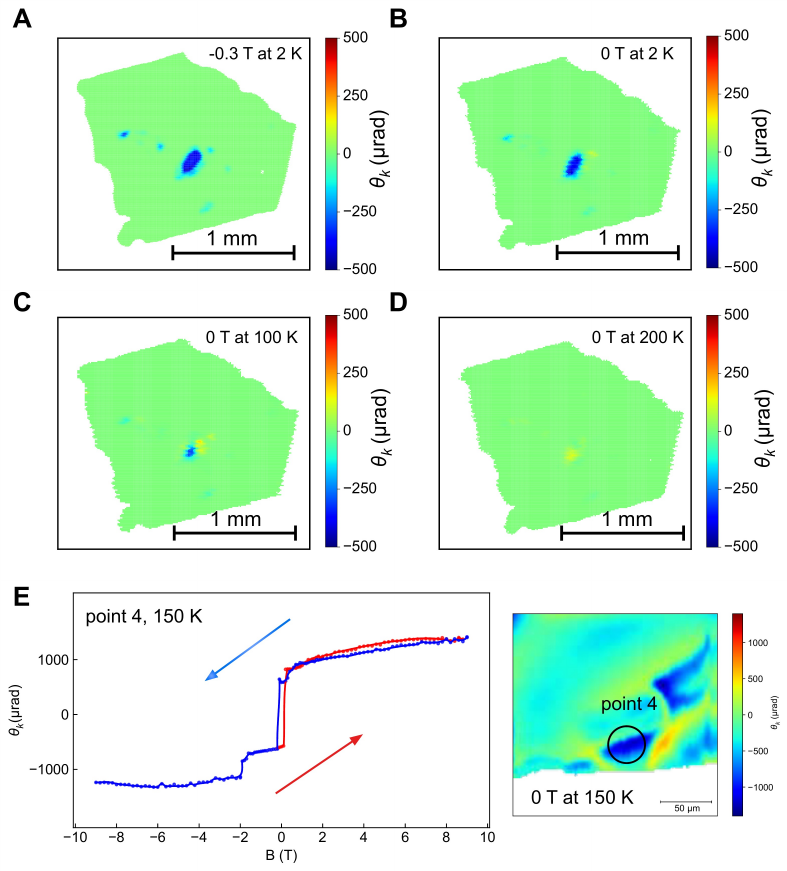}
\caption{\textbf{Domain training via field cooling (FC), temperature-driven chirality
    inversion, and MOKE hysteresis in crystal A.}
    (\textbf{A}) MOKE image at \mbox{$T = 2\,\mathrm{K}$} and \mbox{$B = -0.3\,\mathrm{T}$} after cooling
    from \mbox{$350\,\mathrm{K}$} in a \mbox{$-0.3\,\mathrm{T}$} field.
    (\textbf{B}) Remanent MOKE image at \mbox{$2\,\mathrm{K}$} after field removal, displaying a
    spontaneous signal with negative chirality throughout the Kerr-active regions.
    (\textbf{C}) Persistence of the trained chirality after warming to \mbox{$100\,\mathrm{K}$}.
    (\textbf{D}) Chirality inversion observed after further warming to \mbox{$200\,\mathrm{K}$},
    without any change in applied field.
    (\textbf{E}) Kerr hysteresis measured at point 4 under \mbox{$c$}-axis fields up to
    \mbox{$9\,\mathrm{T}$} at \mbox{$150\,\mathrm{K}$}. The right panel is a zero-field MOKE map of the
    surrounding \mbox{$200\,\mu\mathrm{m}$} region acquired with a \mbox{$1\,\mu\mathrm{m}$} step size.}
\label{fig:3}
\end{figure}

\begin{figure}
\centering
\includegraphics[width=0.95\textwidth]{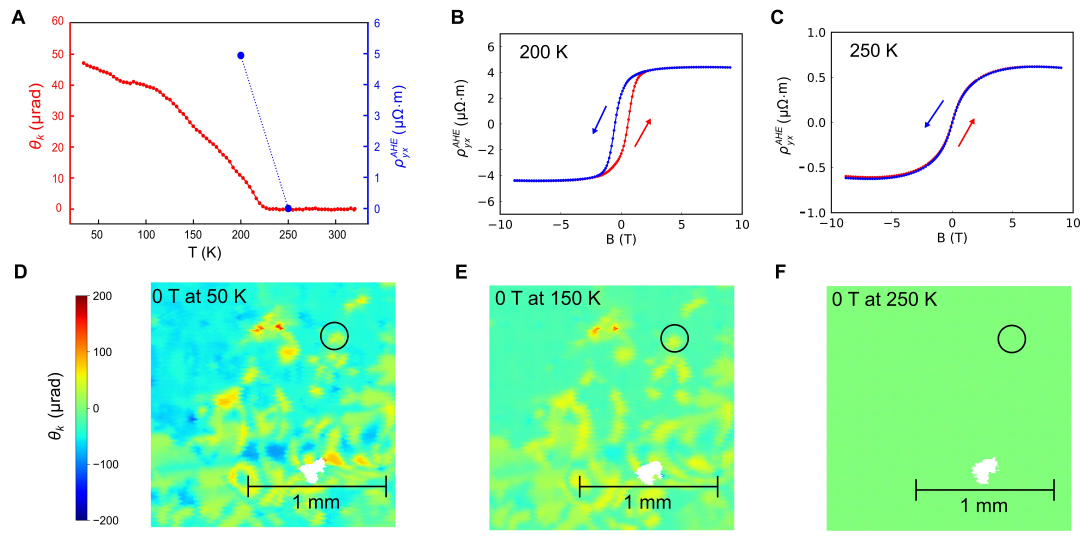}
\caption{\textbf{Spontaneous MOKE and anomalous Hall effect in a conducting
    \mbox{$58\,\mathrm{nm}$} \alphaMnTe{} film.}
    (\textbf{A}) Red: temperature dependence of the spontaneous Kerr rotation \thetaK{} measured
    at the fixed position marked by a black circle in \mbox{(D)--(F).} Blue: anomalous Hall resistivity
    \mbox{$\rho_{yx}^{\mathrm{AHE}}$} extracted from the field sweeps in (B) and (C), plotted on the
    right-hand axis. Both quantities fall to zero above \mbox{${\sim}235\,\mathrm{K}$}.
    (\textbf{B}, \textbf{C}) Anomalous Hall resistivity versus \mbox{$c$}-axis field at
    \mbox{$200\,\mathrm{K}$} and \mbox{$250\,\mathrm{K}$}. Red and blue traces denote increasing and
    decreasing field. At \mbox{$200\,\mathrm{K}$} the loop is clearly hysteretic with a saturated
    amplitude of \mbox{${\sim}4.4\,\mu\Omega{\cdot}\mathrm{m}$}; at \mbox{$250\,\mathrm{K}$} it has closed and
    is reduced by nearly an order of magnitude.
    \mbox{(\textbf{D}--\textbf{F})} Zero-field MOKE maps at \mbox{$50$}, \mbox{$150$}, and \mbox{$250\,\mathrm{K}$} on a
    \mbox{$\pm 200\microrad$} color scale. Positive and negative domains reach
    \mbox{$\pm 200\microrad$} at \mbox{$50\,\mathrm{K}$}, saturating the scale and exceeding the
    \mbox{${\sim}50\microrad$} single-point trace in (A), which was recorded at the black circle rather than
    at a domain maximum. The domains weaken by \mbox{$150\,\mathrm{K}$} and are absent at
    \mbox{$250\,\mathrm{K}$}.}
\label{fig:4}
\end{figure}
\endgroup

\clearpage

\renewcommand{\thefigure}{S\arabic{figure}}
\renewcommand{\thetable}{S\arabic{table}}
\renewcommand{\theequation}{S\arabic{equation}}
\renewcommand{\thepage}{S\arabic{page}}
\setcounter{figure}{0}
\setcounter{table}{0}
\setcounter{equation}{0}
\setcounter{page}{1}

\begin{center}
\section*{Supplementary Materials for\\ \scititle}

Weitung~Yang,
Choongjae~Won,
Cory~D.~Cress,
Weihang~Lu,
Xiaochen~Fang,
Shelby~Fields,
Nicholas~Combs,
Shaofeng~Han,
Marshall~Zachary~Franklin,
Steven~P.~Bennett,
Sang-Wook~Cheong,
Jing~Xia$^{\ast}$\\
\small$^{\ast}$Corresponding author. Email: xia.jing@uci.edu
\end{center}

\subsubsection*{This PDF file includes:}
Supplementary Text\\
Figs. S1 to S12

\newpage

\subsection*{Supplementary Text}

\subsubsection*{Gossamer-ferromagnetism mechanism and the effective $T$-symmetry of ideal
    altermagnetic MnTe}

In the ideal collinear altermagnetic state of \alphaMnTe, the two opposite-spin Mn sublattices
are related by a spin-space two-fold rotation combined with a real-space non-symmorphic six-fold
screw axis
$[C_2 \parallel C_6 t_{1/2}]$~\cite{krempasky2024altermagnetic,amin2024nanoscale}. In the
absence of spin--orbit coupling (SOC), this spin-space rotation leaves the Hamiltonian invariant.
Combined with real-space time reversal, it produces an effective $T$-symmetry—time reversal
combined with a spin flip—that leaves the ideal altermagnetic ground state invariant and forbids
a Berry-curvature contribution to the off-diagonal optical conductivity
\sigmaxy~\cite{unconventional2025moke,smejkal2020crystal}. In the strict nonrelativistic limit,
an ideal altermagnet is therefore magneto-optically inactive at the Berry-curvature level; any
MOKE must arise from the much weaker quantum-metric contribution, which is expected to be below
our current sensitivity~\cite{unconventional2025moke}.

This exact $T$-symmetry is broken in real MnTe by two cumulative effects. First, relativistic
SOC couples the spin-space and real-space symmetries; it has been
emphasized~\cite{cheong2024altermagnetism,cheong2024emergent,cheong2025classification} that real
altermagnets always host some degree of SOC, and the resulting weak ferromagnetism and
TRSB-sensitive responses should be regarded as intrinsic material properties. In MnTe, this SOC
contribution is suppressed to very high order because the leading SOC-induced canting arises only
at third order~\cite{mazin2024origin}, making the pristine MOKE unresolvably small. Second, and more consequentially for the present work, self-doping
introduces itinerant holes into the narrow valence band at the top of the Mn $d$-manifold. In a
strictly collinear antiferromagnetic (AFM) ground state, these holes hop through a zero-bandwidth band and pay a
large kinetic-energy cost; they lower this energy by inducing a small $c$-axis canting $\theta$
of the Mn moments, generating a double-exchange-like kinetic-energy gain that scales as
$\sin(\theta)$, with the canted moment scaling approximately linearly with carrier
density~\cite{mazin2024origin}. This carrier-induced canting breaks the effective $T$-symmetry,
switches on the full Berry-curvature contribution of the altermagnetic bands to \sigmaxy, and
produces a Kerr rotation whose magnitude is set not by the tiny canting angle itself but by the
altermagnetic band structure it unblocks.

\begin{figure}[H]
\centering
\includegraphics[width=0.5\textwidth]{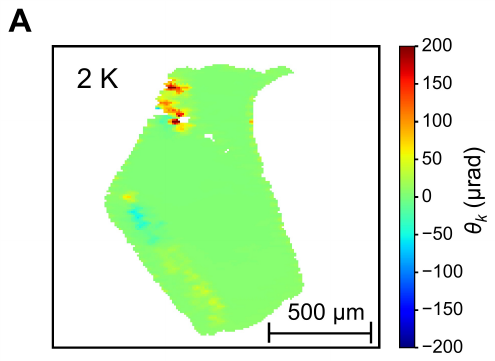}
\caption{\textbf{Reduced spontaneous MOKE in a more resistive bulk crystal C.}
    (\textbf{A}) Zero-field spontaneous MOKE image of bulk single crystal C at \mbox{$2\,\mathrm{K}$}
    after ZFC, showing positive and negative MOKE domains with reduced spontaneous signals of
    \mbox{$\pm 200\microrad$}. Crystal C was grown under nominally identical conditions to
    crystals A and B, but is more resistive, with
    \mbox{$\rhoxx(150\,\mathrm{K}) \approx 100\,\mathrm{m\Omega{\cdot}m}$} versus
    \mbox{${\sim}10\,\mathrm{m\Omega{\cdot}m}$} for crystals A and B. The peak Kerr rotation is
    approximately an order of magnitude smaller than the \mbox{$\pm 1500\microrad$} signal of crystals A
    and B, and four orders of magnitude above the \mbox{$50\nrad$} noise floor.}
\label{fig:S5}
\end{figure}

\begin{figure}[H]
\centering
\includegraphics[width=0.95\textwidth]{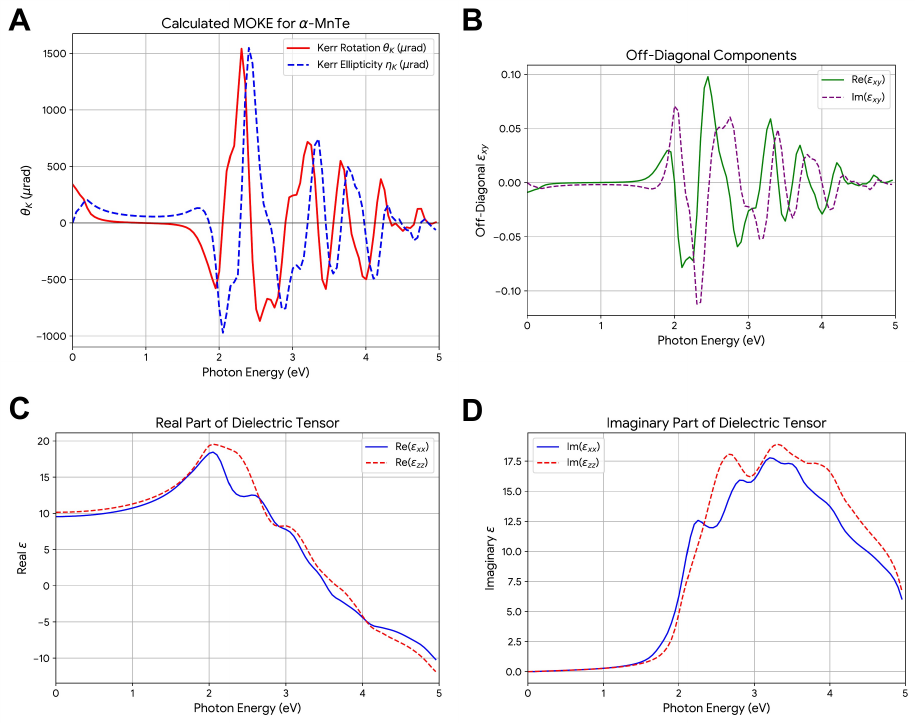}
\caption{\textbf{DFT-calculated MOKE and dielectric tensor of ideal altermagnetic \alphaMnTe.}
    (\textbf{A}) Calculated polar Kerr rotation \thetaK{} (solid red) and Kerr ellipticity
    \etaK{} (dashed blue) versus photon energy, for light incident along the \mbox{$c$}-axis [0001] of
    the ideal altermagnetic state. The calculation is for stoichiometric, undoped \alphaMnTe{}
    and contains neither Mn vacancies nor the resulting carrier-induced canting. The calculated
    Kerr rotation oscillates strongly with photon energy, reaching \mbox{$\pm 1500\microrad$} near
    \mbox{$2.3\,\mathrm{eV}$}, but falls to near zero at the \mbox{$0.80\,\mathrm{eV}$}
    (\mbox{$1550\,\mathrm{nm}$}) energy used in our measurements.
    (\textbf{B}) Real (Re) and imaginary (Im) parts of the off-diagonal dielectric tensor
    component \mbox{$\varepsilon_{xy}(\omega)$}.
    (\textbf{C, D}) Real and imaginary parts of the diagonal dielectric components
    \mbox{$\varepsilon_{xx}(\omega)$} and \mbox{$\varepsilon_{zz}(\omega)$}.
    The calculations are taken from Ref.~\cite{liebman-pelaez_strain_2026}. Because the optical
    calculation is not fully \mbox{$k$}-mesh converged below approximately \mbox{$1\,\mathrm{eV}$}, the
    numerical value at \mbox{$0.80\,\mathrm{eV}$} should be interpreted semiquantitatively; it is
    nevertheless much smaller than the observed bulk response.}
\label{fig:S6}
\end{figure}

This picture accounts for four distinct experimental observations in \alphaMnTe: the capacity of
the altermagnetic band structure to support Kerr rotations of order \mbox{$1000\microrad$} once the
effective \mbox{$T$}-symmetry is broken, even though the undoped calculation predicts a near-vanishing
rotation at our \mbox{$0.80\,\mathrm{eV}$} probe energy (fig.~\ref{fig:S6}); the realization of a
\mbox{$\pm 1500\microrad$} signal at that energy in the hole-conducting bulk crystals A and B; the
reduced signal of \mbox{$\pm 200\microrad$} in the more resistive crystal C (fig.~\ref{fig:S5}); and the
reproduction of the joint MOKE and AHE response in a hole-conducting thin film (Fig.~\ref{fig:4}).
It accounts for the lower edge of the narrow \mbox{$15$}--\mbox{$18\,\mathrm{meV}$}
transport-activation-energy window reported for the AHE~\cite{liu2025strain}, since a canting that
grows with carrier density implies a doping threshold. The upper edge of that window lies outside
the present framework: a monotonic canting predicts no reduction of the response at higher doping,
so an additional mechanism must set the upper bound.

\subsubsection*{Excluding ferromagnetic contamination as the origin of the bulk MOKE}

We considered carefully whether the $\pm 1500\microrad$ signal observed in bulk MnTe crystals
could arise from trace ferromagnetic impurities or secondary phases rather than from the
altermagnetic order. Several lines of evidence rule this out.

\begin{figure}[H]
\centering
\includegraphics[width=0.95\textwidth]{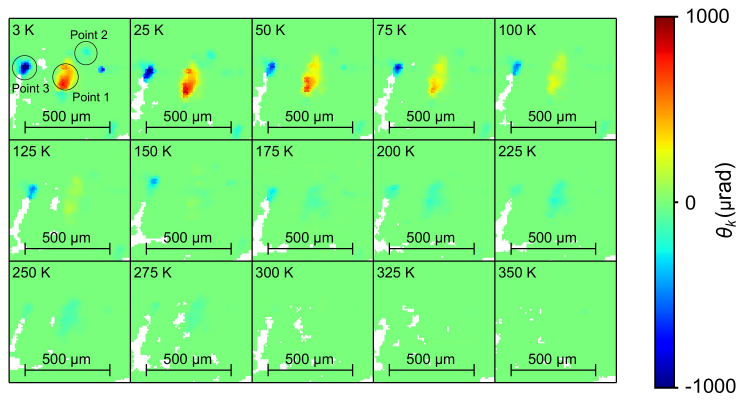}
\caption{\textbf{Spontaneous MOKE domain evolution ($\pm 1000\microrad$ scale).}
    MOKE images of crystal B acquired during ZFW after ZFC, from $3\,\mathrm{K}$ to
    $350\,\mathrm{K}$ in $25\,\mathrm{K}$ steps, plotted on a $\pm 1000\microrad$ color scale.
    Points 1, 2, and 3 used in main-text Fig.~\ref{fig:2}M are marked in the $3\,\mathrm{K}$
    panel. The spontaneous signal nucleates stochastically upon cooling through \TN, grows as the
    altermagnetic order parameter develops, and vanishes on warming across $\TN = 307\,\mathrm{K}$.}
\label{fig:S2}
\end{figure}

\begin{figure}[H]
\centering
\includegraphics[width=0.95\textwidth]{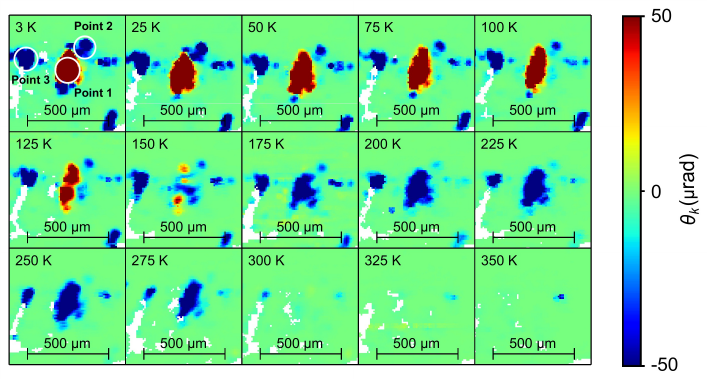}
\caption{\textbf{Detailed MOKE domain evolution (\mbox{$\pm 50\microrad$} scale).}
    Same dataset as fig.~\ref{fig:S2}, plotted on an expanded \mbox{$\pm 50\microrad$} color scale to highlight
    the fine spatial and temperature evolution of the altermagnetic domains. The expanded scale
    reveals additional domains at the \mbox{$10$}--\mbox{$50\microrad$} level that are not visible on the
    \mbox{$\pm 1000\microrad$} scale.}
\label{fig:S3}
\end{figure}

\textit{Onset temperature.} The MOKE signal onsets precisely at \mbox{$\TN = 307\,\mathrm{K}$}
(Fig.~\ref{fig:2}M and figs.~\ref{fig:S2}, \ref{fig:S3}). This does not coincide with the N\'{e}el or Curie
temperatures of any candidate secondary phase: MnO ($\TN \approx 118\,\mathrm{K}$),
MnO$_2$ ($\TN \approx 92\,\mathrm{K}$), \MnthreeOfour{} ($\TC \approx 42\,\mathrm{K}$),
$\beta$-Mn ($\TC < 100\,\mathrm{K}$), or elemental Te (nonmagnetic). No known Mn-containing
secondary phase has a magnetic transition at \mbox{$307\,\mathrm{K}$}.

\textit{Spatial domain structure.} The observed domains are macroscopic
($100\,\mu\mathrm{m}$--$1\,\mathrm{mm}$), containing structure resolved down to the micron
scale (Fig.~\ref{fig:3}E), and take distinct, reproducible morphologies rather
than appearing as isolated dots consistent with impurity clusters. Field cooling reproducibly
trains the chirality, as expected for altermagnetic domains coupled via a weak ferromagnetic
moment to the external field.

\textit{Correlation with intrinsic transport.} The MOKE amplitude tracks an intrinsic transport
property, the sample conductivity, across the bulk crystals: crystals A and B against the more
resistive crystal C. This ordering is consistent with the \mbox{$15$--$18\,\mathrm{meV}$} AHE
activation window~\cite{liu2025strain}. An extrinsic ferromagnetic contamination would not
be expected to correlate systematically with MnTe's own transport activation energy.

\subsubsection*{Comparison of MOKE in MnTe to other spin-compensated systems}

The same zero-loop fiber-optic Sagnac interferometer used in the present work was recently
employed to demonstrate a topological MOKE in the noncoplanar antiferromagnet
\CoTaS~\cite{farhang2026topological}, and to image the Berry curvature
in the coplanar non-collinear antiferromagnet \MnthreeNiN~\cite{farhang2026temperature}. In \CoTaS, a
tetrahedral triple-$Q$ spin configuration generates a non-vanishing scalar spin chirality and a
SOC-free fictitious magnetic field; in \MnthreeSn~\cite{higo2018large} and
\MnthreeNiN~\cite{farhang2026temperature}, coplanar non-collinear triangular spin structures produce large
MOKE via SOC-induced Berry curvature. Both mechanisms yield Kerr rotations of several hundred
\mbox{$\mu\mathrm{rad}$} despite vanishing net magnetization. The same Sagnac platform has also been
used to identify TRSB-breaking superconducting phases, most notably the chiral state of
UPt\mbox{$_3$}~\cite{schemm2014observation}.

\alphaMnTe{} represents a distinct third regime. Its order is \textit{collinear} and
centrosymmetric, with a $g$-wave altermagnetic band structure that in the ideal limit possesses
an effective $T$-symmetry forbidding Berry-curvature MOKE (Supplementary Text). The giant bulk signal is extrinsically activated by carrier-induced canting enabled
by SOC, entirely consistent with the broader
view~\cite{cheong2024altermagnetism,cheong2024emergent,cheong2025classification} that SOC is an
inevitable ingredient of real altermagnets and controls their macroscopic TRSB responses.
\alphaMnTe{} therefore complements the non-collinear and noncoplanar cases and completes the
picture of MOKE mechanisms in spin-compensated magnets.

\subsubsection*{Co-registered optical contrast in the conducting film}

\begin{figure}[H]
\centering
\includegraphics[width=0.95\textwidth]{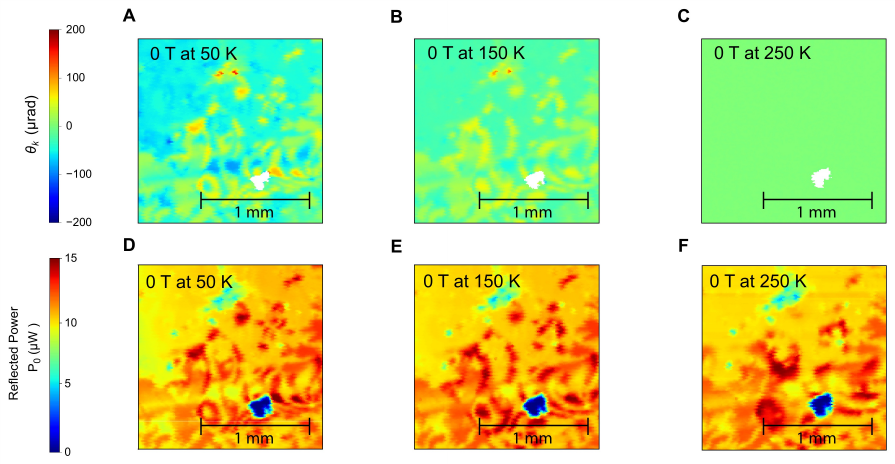}
\caption{\textbf{Co-registered reflected optical power for the conducting film of
    Fig.~\ref{fig:4}.}
    \mbox{(\textbf{A--C})} Zero-field Kerr rotation \thetaK{} of the \mbox{$58\,\mathrm{nm}$}
    \alphaMnTe/InP(111)A film at \mbox{$50$}, \mbox{$150$}, and \mbox{$250\,\mathrm{K}$} on a
    \mbox{$\pm 200\microrad$} color scale, corresponding to \mbox{Fig.~\ref{fig:4}D--F}. The black
    circle marks the fixed position of the \thetaK\mbox{$(T)$} trace of Fig.~\ref{fig:4}A.
    \mbox{(\textbf{D--F})} Reflected optical power \mbox{$P_0$} recorded in the same scans as
    \mbox{(A--C)}, and therefore co-registered with them, on a common
    \mbox{$0$--$15\,\mu\mathrm{W}$} scale. The \mbox{$P_0$} pattern is unchanged from \mbox{$50$} to
    \mbox{$250\,\mathrm{K}$}, over which range the Kerr signal decays to zero. Images were
    registered using the large low-\mbox{$P_0$} feature near the center of the field of view to
    correct for thermal drift between temperatures.}
\label{fig:S10}
\end{figure}

The reflected optical power was recorded simultaneously with the Kerr angle for the
conducting film of Fig.~\ref{fig:4}, as it was for the bulk crystals. Fig.~\ref{fig:S10} shows the
two channels at \mbox{$50$}, \mbox{$150$}, and \mbox{$250\,\mathrm{K}$}. The Kerr maps decay to
zero over this range while the \mbox{$P_0$} pattern, including its fine structure, is unchanged;
the optical contrast of the film is therefore static and not a consequence of the magnetic state,
as in the bulk crystals. Because the images drift in \mbox{$x$}--\mbox{$y$} between temperatures,
they were registered using the large low-\mbox{$P_0$} feature near the center of the field of
view.

We do not interpret the film's optical contrast microscopically, for two reasons specific to
the thin-film geometry. First, the film is measured as grown rather than on a cleaved face, so
local surface topography can modulate the power collected in a confocal geometry. Second, at
\mbox{$0.80\,\mathrm{eV}$} the refractive indices of \alphaMnTe{} and InP are closely matched, so
the film-substrate interface reflects weakly and \mbox{$P_0$} is dominated by the front surface;
the contrast therefore cannot be read as a thickness map either. The film's non-uniform Kerr
pattern is accordingly reported as an observation co-registered with a static optical contrast,
not as independent evidence for the spatial argument, which rests on the bulk crystals.

\subsubsection*{Dependence of the domain pattern on cooling history}

\begin{figure}[H]
\centering
\includegraphics[width=0.95\textwidth]{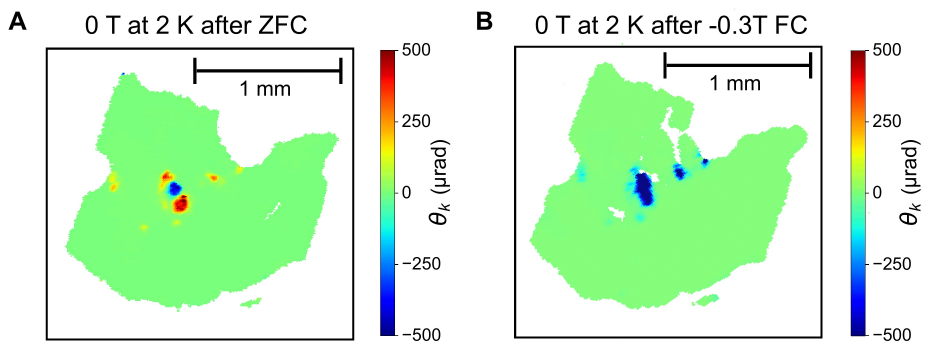}
\caption{\textbf{Dependence of the domain pattern on cooling history in crystal A.}
    (\textbf{A}) Zero-field MOKE image at \mbox{$2\,\mathrm{K}$} after ZFC from above \TN.
    (\textbf{B}) Zero-field MOKE image of the same region at \mbox{$2\,\mathrm{K}$} after FC in
    \mbox{$-0.3\,\mathrm{T}$}, on the same \mbox{$\pm 500\microrad$} color scale.
    The cooling protocol controls the sign of the domains: after ZFC both chiralities appear,
    whereas after \mbox{$-0.3\,\mathrm{T}$} FC all clearly Kerr-active domains in the compared region
    have negative chirality. The same broad Kerr-active locations are retained within the
    registration and spatial resolution, and the same areas remain Kerr-silent in both images.}
\label{fig:S11}
\end{figure}

The MOKE images presented in the main text were acquired during zero-field warming, either after
ZFC (Fig.~\ref{fig:2}) or after FC in \mbox{$-0.3\,\mathrm{T}$} (Fig.~\ref{fig:3}). To separate what
the cooling protocol does control from what it does not, fig.~\ref{fig:S11} compares two
zero-field images of the same region of crystal A at \mbox{$2\,\mathrm{K}$}, one obtained after ZFC and
one after FC in \mbox{$-0.3\,\mathrm{T}$}.

The two images differ in the sign of the Kerr response. After ZFC (fig.~\ref{fig:S11}A) the
region contains both positive and negative domains, with chiralities selected stochastically as
the sample cools through \TN. After FC in \mbox{$-0.3\,\mathrm{T}$} (fig.~\ref{fig:S11}B) all
clearly Kerr-active domains in the compared region have negative chirality, the field having
trained them to a single sense. The
cooling protocol therefore determines the sign of the domains, consistent with the training
experiment of main-text Fig.~\ref{fig:3}.

\begin{figure}[H]
\centering
\includegraphics[width=0.95\textwidth]{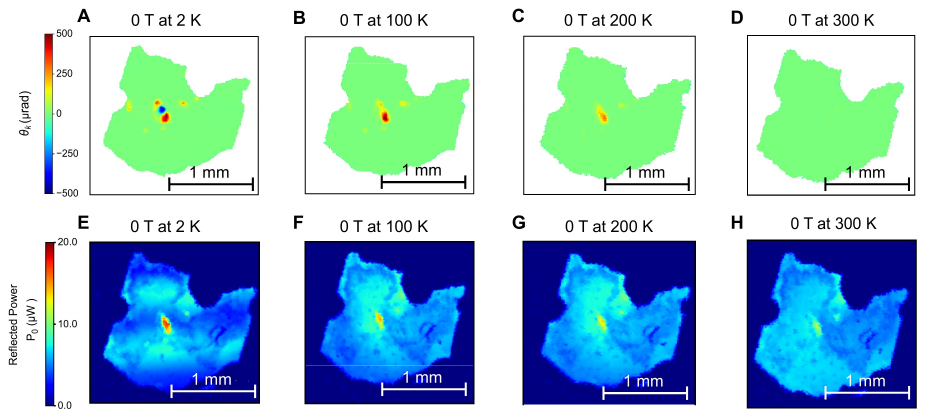}
\caption{\textbf{Co-registered reflected optical power for the ZFC MOKE images of
    Fig.~\ref{fig:2}, shown at larger scale.}
    \mbox{(\textbf{A--D})} Zero-field Kerr rotation \thetaK{} of crystal B acquired during ZFW at
    \mbox{$2$}, \mbox{$100$}, \mbox{$200$}, and \mbox{$300\,\mathrm{K}$} on a
    \mbox{$\pm 500\microrad$} color scale.
    \mbox{(\textbf{E--H})} Reflected optical power \mbox{$P_0$} recorded simultaneously with
    \mbox{(A--D)}, on a common \mbox{$0$--$20\,\mu\mathrm{W}$} scale. Panels reproduce
    \mbox{Fig.~\ref{fig:2}A--H} at a size suitable for inspecting the co-registration.}
\label{fig:S7}
\end{figure}

\begin{figure}[H]
\centering
\includegraphics[width=0.95\textwidth]{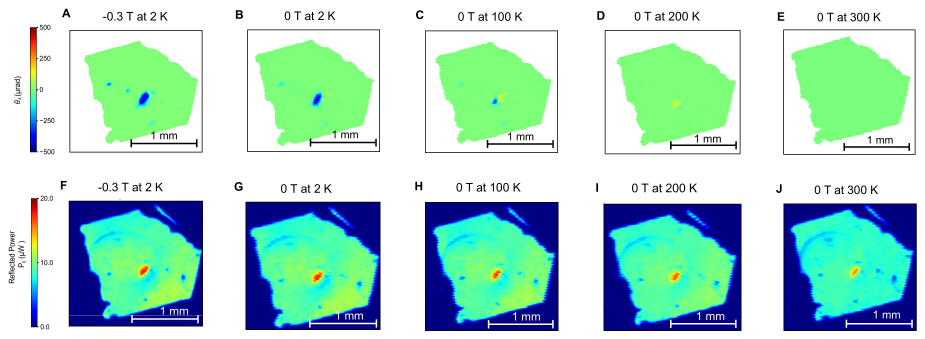}
\caption{\textbf{Co-registered reflected optical power for the field-trained MOKE images of
    Fig.~\ref{fig:3}.}
    \mbox{(\textbf{A--E})} Kerr rotation \thetaK{} of crystal A at \mbox{$2\,\mathrm{K}$} in an applied
    \mbox{$c$}-axis field of \mbox{$-0.3\,\mathrm{T}$} after FC from \mbox{$350\,\mathrm{K}$} (A), and in zero field
    after field removal at \mbox{$2$}, \mbox{$100$}, \mbox{$200$}, and \mbox{$300\,\mathrm{K}$} \mbox{(B--E),} on a
    \mbox{$\pm 500\microrad$} color scale.
    Panels \mbox{(A--D)} correspond to the images of \mbox{Fig.~\ref{fig:3}A--D.}
    \mbox{(\textbf{F--J})} Reflected optical power \mbox{$P_0$} recorded simultaneously with
    \mbox{(A--E),} on a common \mbox{$0$--$20\,\mu\mathrm{W}$} scale. The trained domain coincides
    with a localized maximum of \mbox{${\sim}20\,\mu\mathrm{W}$} against a
    \mbox{$5$--$10\,\mu\mathrm{W}$} background, unchanged by the applied field, by field removal, and
    by warming to \mbox{$300\,\mathrm{K}$} (J) where the Kerr signal is absent (E).}
\label{fig:S8}
\end{figure}

\begin{figure}[H]
\centering
\includegraphics[width=0.75\textwidth]{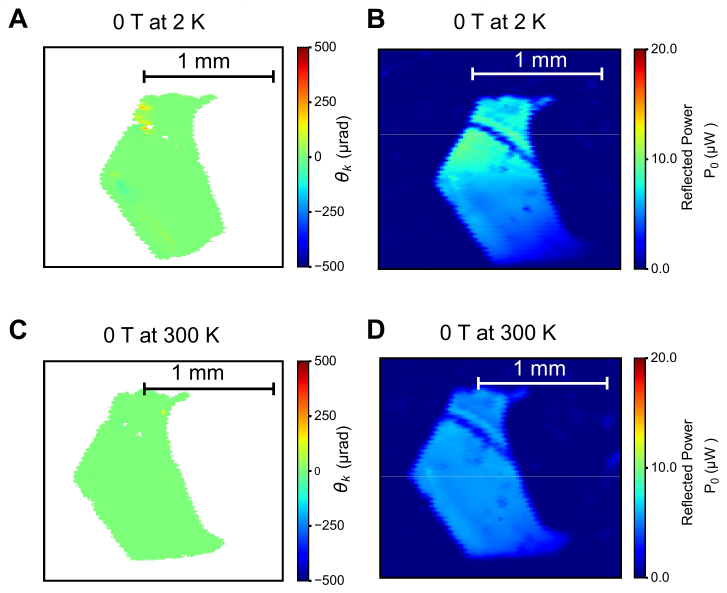}
\caption{\textbf{Co-registered reflected optical power for the more resistive crystal C of
    fig.~\ref{fig:S5}.}
    (\textbf{A}, \textbf{C}) Zero-field Kerr rotation \thetaK{} of crystal C at \mbox{$2\,\mathrm{K}$}
    after ZFC and at \mbox{$300\,\mathrm{K}$}, on the same \mbox{$\pm 500\microrad$} color scale used in
    figs.~\ref{fig:S7} and \ref{fig:S8}.
    (\textbf{B}, \textbf{D}) Reflected optical power \mbox{$P_0$} recorded simultaneously with (A)
    and (C), on the same \mbox{$0$--$20\,\mu\mathrm{W}$} scale. In this more resistive crystal
    \mbox{$P_0$} remains below \mbox{${\sim}10\,\mu\mathrm{W}$} across the face, with no localized
    maxima approaching the \mbox{${\sim}20\,\mu\mathrm{W}$} values of crystals A and B.}
\label{fig:S9}
\end{figure}

The same broad Kerr-active locations are retained within the registration and spatial
resolution. Those locations carry a large Kerr rotation in both images, and the same extended
areas remain Kerr-silent in both, even though the signs at those locations differ. The pattern of \mbox{\textit{where}}
the giant MOKE appears is thus not set by the magnetic history. It instead correlates with a
static optical contrast, independently visible in the co-registered reflection maps of
\mbox{\mbox{figs.~\ref{fig:S7}--\ref{fig:S9}},} that is consistent with spatially varying self-doping.
This separation is what the
carrier-activation picture would predict: local hole doping determines which regions are capable of supporting a
canted moment and hence a Kerr signal, while the cooling field determines the chirality that each
such region adopts.

\begin{figure}[H]
\centering
\includegraphics[width=0.95\textwidth]{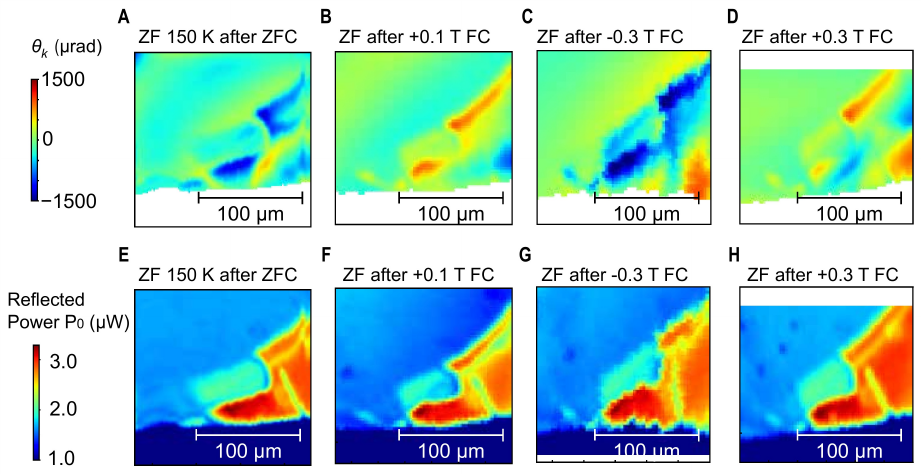}
\caption{\textbf{Dependence of the trained domain state on the cooling field.}
    \mbox{(\textbf{A--D})} Zero-field Kerr rotation \thetaK{} of a \mbox{$200\,\mu\mathrm{m}$}
    region of crystal A at \mbox{$150\,\mathrm{K}$}, the region shown in the right panel of
    Fig.~\ref{fig:3}E, after cooling from \mbox{$350\,\mathrm{K}$} in zero field (A),
    \mbox{$+0.1\,\mathrm{T}$} (B), \mbox{$-0.3\,\mathrm{T}$} (C), and
    \mbox{$+0.3\,\mathrm{T}$} (D), on a \mbox{$\pm 1500\microrad$} color scale. After ZFC both
    chiralities are present; \mbox{$+0.1\,\mathrm{T}$} trains part of the region positive, while
    \mbox{$\pm 0.3\,\mathrm{T}$} train most of the region to the corresponding sign.
    \mbox{(\textbf{E--H})} Reflected optical power \mbox{$P_0$} recorded in the same scans as
    \mbox{(A--D)}, on a common \mbox{$1.0$--$3.2\,\mu\mathrm{W}$} scale. The same broad features are
    retained across the four protocols, although their absolute intensity varies. White pixels
    mark insufficient reflected power.}
\label{fig:S12}
\end{figure}

To examine how the cooling field itself controls the trained state, fig.~\ref{fig:S12}
compares zero-field Kerr images of a single \mbox{$200\,\mu\mathrm{m}$} region of crystal A at
\mbox{$150\,\mathrm{K}$}, the region shown in the right panel of Fig.~\ref{fig:3}E, after cooling
from \mbox{$350\,\mathrm{K}$} in four different fields. After ZFC (fig.~\ref{fig:S12}A) the region
contains both positive and negative domains. After FC in \mbox{$-0.3\,\mathrm{T}$}
(fig.~\ref{fig:S12}C) most of the region is trained to negative chirality, and after FC in
\mbox{$+0.3\,\mathrm{T}$} (fig.~\ref{fig:S12}D) most of it is trained to positive chirality, the
opposite-sign counterpart of the negative-field result. A smaller \mbox{$+0.1\,\mathrm{T}$} field
(fig.~\ref{fig:S12}B) trains only part of the region.

Two points follow. Cooling fields of equal magnitude and opposite sign produce trained states
of opposite chirality over the same region, so the training is a genuine response to the field
rather than a fixed property of the region. The extent of the training also grows with the
magnitude of the cooling field: \mbox{$+0.1\,\mathrm{T}$} trains part of the region while
\mbox{$\pm 0.3\,\mathrm{T}$} train most of it, so the training is progressive over this range and
is not yet complete at the largest field used here. The co-registered reflection maps
(\mbox{figs.~\ref{fig:S12}E--H}) retain the same broad features across the four protocols,
although their absolute intensity varies, supporting the conclusion that the optical contrast is
not set by the magnetic history.

\begin{figure}[H]
\centering
\includegraphics[width=0.95\textwidth]{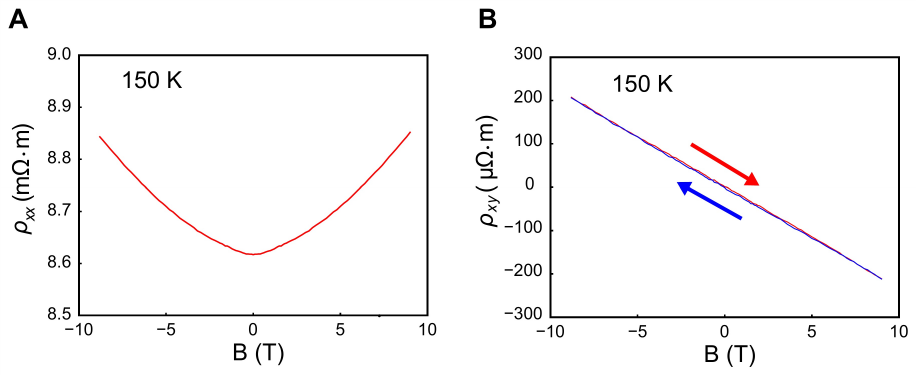}
\caption{\textbf{Hall effect measurements at 150~K.}
    (\textbf{A}) Longitudinal magnetoresistivity \mbox{$\rho_{xx}(B)$} of a MnTe bulk crystal (crystal
    A) at \mbox{$150\,\mathrm{K}$}, showing a weak positive magnetoresistance characteristic of the
    hole-self-doped bulk regime.
    (\textbf{B}) Hall resistivity \mbox{$\rho_{xy}(B)$}, comprising both the linear ordinary Hall
    effect (dominant background) and the AHE hysteresis loop (small \mbox{$\pm 5\,\mu\Omega{\cdot}\mathrm{m}$}
    deviation from linearity). The raw data are plotted as \mbox{$\rho_{xy}(B)$}; the antisymmetrized
    anomalous component is reported in main-text Fig.~\ref{fig:1}C as \rhoAHE{}, following the
    convention \mbox{$\rho_{yx} = -\rho_{xy}$} for the antisymmetric Hall response.}
\label{fig:S1}
\end{figure}

\begin{figure}[H]
\centering
\includegraphics[width=0.95\textwidth]{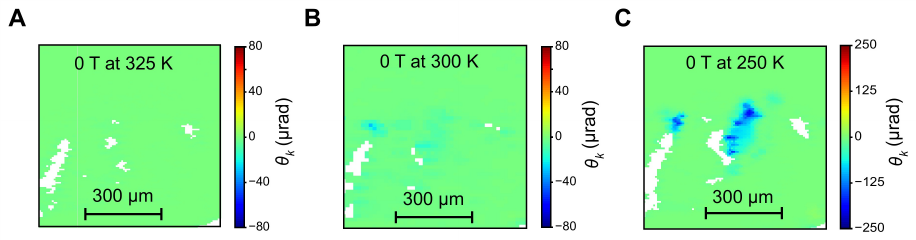}
\caption{\textbf{Temperature dependence of the training effect.}
    (\textbf{A}) Zero-field MOKE image of crystal A at $325\,\mathrm{K}$ (above \TN) after
    removing a $-0.3\,\mathrm{T}$ field; no MOKE signal is observed, confirming that field
    cooling has no effect in the paramagnetic phase.
    (\textbf{B, C}) Zero-field MOKE images at $300\,\mathrm{K}$ and $250\,\mathrm{K}$,
    respectively, after the same field-removal protocol, demonstrating that the negative chirality
    is successfully trained only below \TN. These controls establish that the trained domain state
    reported in main-text Fig.~\ref{fig:3} is a genuine below-\TN{} altermagnetic effect.}
\label{fig:S4}
\end{figure}

\end{document}